\documentclass[submission, Phys]{SciPost}
\usepackage{amsmath,amssymb,amsthm,mathrsfs,graphicx,float,colordvi,url,wrapfig,color}
\usepackage{enumerate,enumitem,here,mathrsfs,wrapfig,ulem,comment}   

\def \be {\begin{equation}} 
\def \ee {\end{equation}}
\def \bea {\begin{eqnarray}}      
\def \eea {\end{eqnarray}}

\begin{document}

\begin{center}{\Large \textbf{ 
\vspace{-1.0mm}
{Hawking flux of 4D Schwarzschild black hole 
\\
\vspace{1.5mm}
with supertransition correction to second-order}
}}\end{center} 

\begin{center}
Shingo Takeuchi
\end{center}

\begin{center}\vspace{-0.5mm} 
Phenikaa Institute for Advanced Study and Faculty of Basic Science, \\
Phenikaa University, Hanoi 100000, Vietnam
\end{center}
 
\vspace{-10mm}
\section*{Abstract}
\vspace{-3.0mm}
{
Former part of this article is the proceedings for my talk \cite{confhomp}  on \cite{Lin:2020gva}, 
which is a report on the issue in the title of this article.    
Later part is the detailed description of \cite{Lin:2020gva}.  
}

\vspace{3.0mm}  

\noindent\rule{\textwidth}{1pt} 
\tableofcontents\thispagestyle{fancy}
\noindent\rule{\textwidth}{1pt}

\allowdisplaybreaks

\section{Motivation for my analyzing Hawking flux}
\label{sec:intro}

We will start with a 4D spacetime, then impose some falloff condition to its spatially infinite region\footnote{  
Talk slide is in the homepage \cite{confhomp}.},  
which means the spacetime asymptotes to the flat spacetime according to that. 
Then, we will consider the diffeomorphism at its spatially infinite region in the range of the falloff condition   
(at this time, gauge conditions to eliminate local diffeomorphism ambiguities are also preserved). 
Transformation of the spacetime by this diffeomorphism is called asymptotic  (or BMS) symmetry \cite{Bondi:1962px,Sachs:1962wk}. 
I list several facts known for this. 
\begin{itemize}
\item   
Diffeomorphism of the asymptotic symmetry is  an infinite dimensional group, 
which contains the Poincar\`{e} group as its subgroup. 
\item    
There are so-called supertranslation and superrotation in the asymptotic symmetry, 
which respectively contains the global translation and Lorentz transformation. 
\item     
Diffeomorphism of the asymptotic symmetry maps 
a configuration of an asymptotically flat spacetime as a solution 
to an other physically different asymptotically flat spacetime as a solution 
in the range of the falloff condition.  
\item 
Asymptotically flat spacetime is infinitely degenerated  
in the range of asymptotic symmetry, 
and the symmetry of theory 
at that asymptotic region is not  Poincar\`{e} symmetry 
but the one  associated with supertranslation and superrotation. 
\item   
Infinite number of conserved charges for supertranslation and superrotation can be defined respectively. 
These can respectively generate  these diffeomorphism, 
however these are given by 2 parts: so-called hard- and soft-parts, which will become creation and annihilation operators for soft-gravitons at quantum level.  
For this, the quantum state for the infinite far region of the asymptotically flat spacetime becomes another one of asymptotically flat spacetime 
when these charges act on these, and soft-gravitons can be considered as a kind of Goldstone boson. 
In this sense, the asymptotic symmetry is some kinds of spontaneously broken symmetry \cite{Strominger:2017zoo}.

\item
Supertranslated spacetimes are normal  \cite{Compere:2016hzt},  therefore considering asymptotic symmetry is meaningful realistically. 
\end{itemize}

In \cite{Lin:2020gva}, I obtain the Schwarzschild black hole spacetime with supertranslation correction to the second-order, 
which I sketch as 
\begin{eqnarray}\label{yjtyt}
ds^2 
\!\!\!&=&\!\!\!
-(1-{2m}/{r} +\cdots+{\cal O}(\varepsilon^3))\,dt_s^2 
+((1-{2m}/{r})^{-1}+\cdots+{\cal O}(\varepsilon^3))\,dr_s^2
\nonumber\\ 
&&
+(r^2+\cdots+{\cal O}(\varepsilon^3))\,d\theta_s^2 
+(r^2\sin^2\theta+\cdots+{\cal O}(\varepsilon^3))\,d\phi_s^2 
\nonumber\\ 
&&
+2((\cdots)\varepsilon+\cdots+{\cal O}(\varepsilon^3)) \,dr_s d \theta_s,
\end{eqnarray}
where $\varepsilon$ mean the order of supertranslation correction 
and the full expressions of metrices are given $j^{(2)}_{\mu\nu}$ in (\ref{ctks001})-(\ref{ctks332}). 
We can obtain the position of the horizon from this as 
\begin{eqnarray}\label{jdjut}
r_{h,4D}  =2 m-\frac{15  m \sin ^2(2 \theta )}{8 \pi } \varepsilon^2+O(\varepsilon ^3).
\end{eqnarray}
The correction of supertranslation enters from the second-order, 
which is the motivation for our analysis to the second-order. 
Here, as this is not constant, there may be a concern for the zeroth law of the black hole thermodynamics.  
It is no problem since the Hawking temperature is constant 
in the range of our analysis's order,  $\varepsilon^2$, as shown below. 

As for the problem with Birkhoff's theorem, 
considering the point that the information is encoded in the asymptotic region, 
it is inferred that the deformed near-horizon geometry is meta-stable and will get settled down to just a Schwarzschild finally, 
while that in the asymptotic region is the stable.

Since the position of the horizon is displaced for supertranslation correction,  
it is interesting to check how the Hawking temperature is. 
Its result is
\begin{eqnarray}\label{hkofewd}
T_H = {1}/{8\pi m}+O(\varepsilon ^3),
\end{eqnarray}
which is no difference from just the Schwarzschild  (reason is written in Sec.\ref{subsec:hwkkdk}).
                                               
Since Hawking temperature can be calculated from Hawking flux,  
if the Hawking temperature were preserved the Hawking flux would be expected to be preserved.  
However I have considered a possibility that supertranslation corrections may be involved in the Hawking flux 
but would be canceled out in the Hawking temperature. 
This is one of my motivations for my computing the Hawking flux in \cite{Lin:2020gva}.  

There is another motivation, which is that 
as a result of involving the supertranslation corrections as in (\ref{yjtyt}), 
it becomes obscure whether field theories can reduce to free 2D or not in the near-horizon.    
Originally it should be so for the strongly gravitational force at the horizon, 
and if not, it would be physically abnormal. 
Although it can be shown in Sec.\ref{subsec:emnh} that the scalar theory can reduce to free 2D,  
whether it is possible or not is unclear before trying  
(I comment on the key for the feasibility of this in the last of Sec.\ref{subsec:emnh}).\newline

There would be many works analyzing the Hawking radiation in some supertranslated situations. 
In these, as the works relating with this study, I take \cite{Chu:2018tzu,Javadinazhed:2018mle,Maitra:2019eix,Wen:2021ahw} 
in the range I know.  From the research situation  mentioned in the following, the analysis in this study would be worthwhile.  

In \cite{Chu:2018tzu}, 
considering Vaidya spacetime with linear order supertranslation correction,   
it is shown that Hawking radiation depends on supertranslation correction 
if the mass depends on the advanced time $v$.
Conversely, if the mass is a constant, $M'=0$, 
there is no correction in Hawking radiation.  
This is consistent with the result in this study, however my analysis is performed to second-order. 

Result in \cite{Javadinazhed:2018mle} is that 
Hawking radiation in the asymptotically flat spacetime given by Bondi coordinates does not get corrected, 
however the position of the horizon in their analysis is assumed to be $2GM$.  
The analysis in this study is performed taking the correction to the position of the horizon into account. 

\cite{Maitra:2019eix} performs linear order analysis 
regarding general diffeomorphisms including supertranslation expressed by $F$ (see (7)), 
then it is concluded 
the proportional coefficient in the relation between the surface gravity and Hawking temperature, 
$1/ 2\pi$ in $T_H = \kappa / 2\pi$, gets some correction (see 2 points: 
``However, macroscopic $\cdots$ transformation.'' in P.2 
and ``as expected $\cdots$ and Hawking [68].'' in P.9, and  (54)). 
This is against to the conservative results in \cite{Chu:2018tzu}, \cite{Javadinazhed:2018mle} and this study. 

In \cite{Wen:2021ahw}, a Hawking radiation in a 4D supertranslated Schwarzschild black hole is analyzed by some expansion around large mass, 
and its result depends on the angular. 
This interferes with the zeroth law of the black hole thermodynamics, and does not agree to \cite{Chu:2018tzu,Javadinazhed:2018mle,Maitra:2019eix} and this study. 
However, 
remember it is the laws for the stable solutions 
and the shape of the asymptotic region is the key in the information paradox. 
Then, the interference with the black hole thermodynamics would not be any problems
if the deformation of the near-horizon geometry is meta-stable, 
while that in the asymptotic region is stable (as mentioned between (\ref{jdjut}) and (\ref{hkofewd})).  
As for the disagreement of results, we could not say anything immediately, since the kind of the expansion is different and cannot compare.

\section{What's supertranslation and its NG boson fields}
\label{sec:intro:wsst} 

We start with an expression of general 4D spacetime 
by the Bondi coordinates $(u,r,\Theta^A)$  ($u=t-r$ and $\Theta^A$ are the  spherical coordinates $(z,\bar{z})$ on the $S^2$) as
\begin{eqnarray}\label{nbop}
ds^2= 
-Udu^2-e^{2\beta}dudr
+g_{AB}
( d\Theta^A +\frac{1}{2}U^A du )
( d\Theta^B +\frac{1}{2}U^B du ),
\end{eqnarray}
where the Bondi gauge is imposed to fix the local diffeomorphisms, which is $g_{rr}=0$, $g_{rA}=0$ and $\partial_r \det (g_{AB}/r^2)=0$. 
Then, supposing that the spacetime will asymptote to the flat spacetime, 
let us consider to describe the neighborhood of ${\mathcal T}^+$. 
At this time we need to impose a falloff condition to the metrices, 
however there is no systematic ways to determine the falloff condition, and various falloff conditions can be considered.  
Typically, it is chosen so that physical solutions can exist and unphysical solutions do not exist. 

As an expansion of  (\ref{nbop}) to $r^{-1}$,
the following one is conventionally adopted \cite{Bondi:1962px,Sachs:1962wk}: 
\begin{align}\label{plfgd}
ds^2   
= & 
- du^2 - 2 du dr + 2r^2 \gamma_{z \bar{z}} d z d \bar{z} 
\nonumber \\
& 
+ {2 m_B}/{r} \,du^2 + r C_{zz} dz^2 +  r C_{\bar{z} \bar{z}} d\bar{z}^2 + D^z C_{zz} du dz 
\nonumber \\
& 
+ \frac{1}{r} \,( \frac{4}{3} ( N_z + u \partial_z m_B ) - \frac{1}{4} (C_{zz} C^{zz} ) )  du d z + \text{c.c.}  +O(r^{-2})\,,
\end{align}
where  $D_z$ is the covariant derivative with respect to $\gamma_{\bar{z}z}$. 
It is usual that the structure to $r^{-1}$ is important.
$C_{zz}$, $C_{\bar{z}\bar{z}}$, $m_B$ and $N_z$ are functions of $(u, z, \bar{z})$ but not of $r$, and
\begin{itemize}
\item 
$m_B$ is the Bondi mass aspect. 
$\int_{S^2} dz d\bar{z} m_B$ gives the Bondi mass, which can be ADM mass in the cases of black hole spacetimes. 
\item 
$N_z$ is the angular momentum aspect. 
$\int_{S^2} dz d\bar{z} N_zV^z$ gives the total angular momentum, which is ADM angular momentum in the black hole spacetimes.  
\item 
$C_{zz}$ and $C_{\bar{z}\bar{z}}$ play the role of potential for gravitational wave 
(akin to vector potential for electromagnetic field), and 
$N_{zz}$ is the Bondi news given as $\partial_u C_{zz}$ ($N_{\bar{z}\bar{z}}$ is likewise).  
\end{itemize}
The falloffs of the metrices in (\ref{plfgd}) are given as follows:
\begin{equation}\label{irtho}
\begin{split}
g_{uu}=&~-1+{\cal O}(r^{-1}), \quad 
g_{ur}=-1+{\cal O}(r^{-2}), \quad 
g_{uz}={\cal O}(1), \\ 
g_{zz}=&~{\cal O}(r), \quad 
g_{z\bar{z}}=r^2\gamma_{z\bar{z}}+{\cal O}(1), \quad 
g_{rr}=g_{rz}=0. 
\end{split}
\end{equation}

Let us turn to the supertranslation. Displacement of metrices,  $m_B$, $N_{zz}$  and $C_{zz}$ in (\ref{plfgd})  
by the diffeomorphism of the supertranslation  is given by the Lie derivatives as
\begin{align}\label{hrtsk}
{\cal L}_\xi  g_{ur}         &= - \partial_u \zeta^u +  {\cal O}  ( r^{-1} ), \nonumber\\
{\cal L}_\xi  g_{zr}         &=  r^2 \gamma_{z\bar{z}} \partial_r \zeta^{\bar{z}} - \partial_z \zeta^u + {\cal O} ( r^{-1} ),  \nonumber\\
{\cal L}_\xi  g_{z\bar{z}} &= r \gamma_{z\bar{z}} ( 2 \zeta^r + r D_z \zeta^z + r D_{\bar{z}} \zeta^{\bar{z}} ) + {\cal O} (1), \nonumber\\
{\cal L}_\xi  g_{uu}        &= - 2 \partial_u \zeta^u - 2 \partial_u \zeta^r + {\cal O} ( r^{-1}), \\[0.5mm]
{\cal L}_\xi  m_B &= f \partial_u m_B + \frac{1}{4} ( N^{zz} D_z^2 f + 2 D_z N^{zz}  D_z f + c.c. ), \nonumber\\
{\cal L}_\xi  N_{zz} &= f \partial_u N_{zz}, \nonumber\\
{\cal L}_\xi  C_{zz} &= f \partial_u C_{zz} - 2 D_z^2 f, \nonumber
\end{align}
where the vector field proscribing the coordinate transformation in the Lie derivatives above is given as
\begin{align}\label{nlfbs}
\xi 
= 
f \partial_u 
+ \frac{1}{r} ( D^z f \partial_z + D^{\bar{z}} f \partial_{\bar{z}} ) 
+ D^z D_z f\partial_r.
\end{align}
$f$ is arbitrary function of $(z,\bar{z})$, and normally spherical harmonics are taken. The field referred to as NG boson field is defined for one for every $f$  as
\begin{align}\label{rgese}
{\cal L}_\xi  C(z,\bar{z}) =f(z,\bar{z}).   
\end{align}

\section{Fun in the asymptotic symmetry}
\label{sec:fu}

First of all, 
what 4D Minkowski spacetime has not been 
an unique vacuum 
but infinitely degenerated 
would be a surprisingly interesting fact. 
This had been already found in 1962 \cite{Bondi:1962px,Sachs:1962wk}, 
however it is in just the last decade that hep-th has recognized this problem \cite{Barnich:2009se,Hawking:2016msc}. 
As interesting directions from the study of the asymptotic symmetry, the following ones could be taken: 
{\bf 1)} gravitational memory effect,  
{\bf 2)} link with soft theorems  and holography, 
and  
{\bf 3)} information paradox.  
\newline

{\bf 1)} is the variation in the relativistic position of two objects near the future null infinity ${\cal T^+}$ 
for the passing of the gravitational wave, 
which could be measured by the formalism of the asymptotic symmetry.

Consider the gravitational wave is turned on at $u=u_i$ and off at $u=u_f$, 
and two objects near ${\cal T^+}$ are exposed it during the time interval $\Delta u=u_f-u_i$.  
The Bondi news tensor and the energy momentum tensors are zero at any time except for the time getting the gravitational wave.
Then, one can evaluate the displaced amount as
\begin{align}\label{klwef} 
\Delta s^{\bar{z}}={\gamma^{z\bar{z}} \over 2r} \Delta C_{zz}s^z, 
\end{align} 
where $\Delta s^A=s^A|_{u=u_f}- s^A|_{u=u_i}$ 
($s^A$ mean the relativistic position of the two objects),  
and $\Delta C_{AB}=C_{AB}|_{u=u_f}-C_{AB}|_{u=u_i}$.   
  
Hence, the passing of gravitational wave is considered to arise the displacement by the order $r^{-1}$. 
Now observation of the gravitational memory effect is undergoing \cite{Pshirkov:2009ak,vanHaasteren:2009fy,NSeto,Wang:2014zls,Arzoumanian:2015cxr,Lasky:2016knh}.

Other types of memory effect are also considered:  
spin memory effect \cite{Pasterski:2015tva}, 
color memory effect \cite{Pate:2017vwa}, 
and electromagnetic memory effect \cite{Bieri:2013hqa,Susskind:2015hpa,Pasterski:2015zua}. 
Observing the soft graviton may be also planed, 
however it is so silent that it is not caught in our current detection.
\newline
  
Regarding {\bf 2)}, the equation of the soft theorem can be obtained  
from the Ward identity with regard to the asymptotic symmetry 
(\hspace{-0.5pt}\cite{Strominger:2013lka,He:2014cra} and \cite{Strominger:2013jfa,He:2014laa,Cachazo:2014fwa,Kapec:2014opa,Kapec:2016jld,He:2017fsb}  
for gauge theories and gravity, respectively). 
Therefore,   
\begin{align}\label{klgbe}
\textrm{asymptotic symmetry} \,\longleftrightarrow\, \textrm{soft theorem.}  
\end{align}

Currently the correspondence between the $S$-matrix in 4D asymptotically flat spacetime and 2D CFT are ongoing 
\cite{He:2015zea,Bagchi:2016bcd,Pasterski:2016qvg,Cardona:2017keg,Pasterski:2017kqt,Pasterski:2017ylz}.

Next, the DC shift (equation given in P.91 of \cite{Strominger:2017zoo}) 
and ``the effect of attaching one soft-graviton line to an arbitrary Feynman diagrams'' 
can be identical each other via Fourier transformation (with adjustment of some notation’s conventions). 
From this fact, it is considered that 
the gravitational wave from black holes 
and the soft particles from the elementary particle’s collisions 
will show analogous behavior at the long distance in the observation \cite{Strominger:2017zoo}.
Thus, as the phenomena showing analogous behavior \cite{Strominger:2014pwa},
\begin{align}\label{lrgfdd}
\textrm{soft theorem} \,\longleftrightarrow\, \textrm{memory effect}. 
\end{align}

Lastly, the gravitational wave at the long distance can be considered 
as a kind of diffeomorphism of the asymptotic symmetry, and has a relation with memory effect. Hence, 
\begin{align}\label{sdgavs}
\textrm{memory effect} \,\longleftrightarrow\, \textrm{asymptotic symmetry}. 
\end{align}

It is very interesting that different theories and phenomena can get related like the one above. 
Same relations can be obtained in the gauge field theories \cite{Strominger:2017zoo}.  
Therefore, gravitational and gauge theories would be universal in the IR-region.  
\newline 

Regarding {\bf 3)}, an initial configuration to form a star or black hole finally leads to some deformed spacetimes by supertranslation 
(for an explicit analysis for this, see \cite{Compere:2016hzt}), and its phase space is infinite dimensional. 
Hence, we can expect that the information of the initial configuration could be preserved in the final shape of the spacetime, 
which is the scenario we can highly expect as the solution to the information paradox \cite{Hawking:2016msc,Hawking:2016sgy,Strominger:2017aeh}.

\section{Our 2D effective action with supertranslation correction} 
\label{sec:2d}

From here, I would like to talk on my study. 
What I want to do first is to obtain the Schwarzschild black hole metric 
with the supertranslation correction to the second order 
in the Schwarzschild black hole coordinates.  
For this we will start with the Schwarzschild black hole spacetime given in 
the isotropic coordinates: 
\begin{eqnarray}\label{vpoe}
ds^2= 
-\frac{( 1 - {m}/{2\rho_s} )^2}{( 1 + {m}/{2\rho_s} )^2}dt_s^2
+( 1 + {m}/{2\rho_s} )^4 ( d\rho_s^2+\rho_s^2 d\Omega_s^2 ), 
\end{eqnarray}
where a flat three-dimensional space part, $d\rho_s^2+\rho_s^2 d\Omega_s^2 $, 
is convenient to involve the supertranslation correction 
according to \cite{Compere:2016hzt}. 
\newline
 
Then, writing as $d\rho_s^2+\rho_s^2 d\Omega_s^2= dx_s^2 + dy_s^2+dz_s^2$
and $\rho_s^2 =  x_s^2+y_s^2+z_s^2$, we involve the supertranslation correction according to \cite{Compere:2016hzt}:
\begin{subequations}
\label{ctki0}
\begin{align} 
\label{ctki1}
x_s &= ( \rho-C ) \sin\theta \cos\phi+\sin \phi \csc \theta \,\partial_\phi C-\cos\theta\cos\phi \, \partial_\theta C,
\\
\label{ctki2}
y_s &= ( \rho-C ) \sin\theta \sin\phi- \cos \phi \csc \theta \,\partial_\phi C-\cos\theta\sin\phi \, \partial_\theta C,
\\
\label{ctki3}
z_s &= ( \rho-C ) \cos\theta+\cos\theta\cos\phi \, \partial_\theta C, 
\end{align}
\end{subequations}
where the function $C$ is the NG boson field for supertranslation, which we will take as 
\begin{eqnarray}\label{cdsvds}
C = m\varepsilon\, Y_2^0(\theta,\phi). 
\end{eqnarray}  
\begin{itemize} 
\item 
$\varepsilon$ is dimensionless, 
which we attach to measure the order of supertranslations in our analysis. 
$m$ is that in (\ref{vpoe}), 
which we involve to have $C$ have the same dimension with $\rho$ (now, $G/c^2=1$). 
The correction of $\varepsilon$  appears from the second-order (see (\ref{jdjut})) in the position of the horizon, 
which is our motivation for the analysis to $\varepsilon^2$-order.  

\item 
Why we consider $Y_2^0$ that 
this mode is expected to be dominant 
in the process forming a soft-hairy black hole (e.g. \cite{Berti:2005ys}). 
We have also performed the analysis with $Y_1^0$ just in case. 
Although we have not performed the calculation to the end, 
it has been seemed to be  essentially same with what will present in the following. 
\end{itemize}

Involving  (\ref{ctki0}) into  the isotropic coordinates (\ref{vpoe})  to $\varepsilon^2$-order, 
we will rewrite it into the Schwarzschild coordinates (for detail, see Sec.\ref{sec:mgwc}), 
and finally obtain like (\ref{yjtyt}). 
\newline

Then, with these 4D metrics, we consider a complex scalar field theory as
\begin{align}\label{bkjdsa}
S=\int d^4x \sqrt{-g}\,g^{MN}\partial_M \phi^* \partial_N \phi.
\end{align}
Writing $\phi(t,r,\theta, \phi)=\phi_{lm}(t,r)Y_m^l(\theta, \phi)$, 
and taking near-horizon limit by writing $r=r_{h,4D}+\Delta r$, 
we can get the 4D near-horizon  action as 
\begin{subequations}
\label{vnksld}
\begin{align}
\label{vnksld1}
S   
=&  
-\sum_{l, m} \sum_{k, n} \int dt dr (2m)^2  
\Big\{
\nonumber\\*
&
\hspace{5.0mm}
\phi_{lm}^* 
\Big( 
-\frac{2m}{r-2m}\Lambda_{lm,\,kn} 
-\frac{15m^2\varepsilon^2}{4\pi(r-2m)^2}
\! \int \!  
d\Omega \sin^2 (2\theta)
(Y_l^m)^* Y_k^n
\Big) 
\partial_t \partial_t \phi_{kn}
\nonumber\\*
&
+\phi_{lm}^* \partial_r 
\Big(
\frac{r-2m}{2m}\Lambda_{lm,\,kn} 
-\frac{15m^2\varepsilon^2}{16\pi}
\! \int \! d\Omega\sin^2 (2\theta)
(Y_l^m)^* Y_k^n
\Big)  
\partial_r\phi_{kn}
\Big\}
\!+O(\varepsilon ^3)\,,
\nonumber\\* 
\\
\label{vnksld2}
\Lambda_{lm,\,kn} 
\equiv& 
\int d\Omega
\Big\{
1+\frac{3}{2} \sqrt{\frac{5}{\pi }} \left(1 +3 \cos (2 \theta ) \right) \varepsilon 
\nonumber\\* 
&
+\frac{45}{2\pi}  
\Big(
\frac{ \sin (2 \theta)}{4} -\cos ^2(\theta )+3  \cos ^2(\theta ) \cos (2 \theta ) 
\Big) \varepsilon ^2
\Big\} \, (Y_l^m)^*Y_k^n. 
\end{align}
\end{subequations} 

Integrating out ($\theta$, $\phi$), we obtain 2D near-horizon effective action as 
(Sec.\ref{subsec:nhar} and \ref{subsec:emnh})
\begin{align}\label{bqws}
S_{\textrm{2D eff}}= 
\sum_{l=0}^{l_{max}} \sum_{|m|=0}^l 
\int d^2x \, \Phi_{lm}
\big( 
(g_{\rm eff})^{tt}_{lm} \, \partial_t \varphi^*_{lm}  \partial_t  \varphi_{lm}
+(g_{\rm eff})^{rr}_{lm} \, \partial_r  \varphi^*_{lm}  \partial_r  \varphi_{lm}
\big),
\end{align}
\vspace{-5mm} 
\begin{subequations}
\label{powmn}
\begin{align}
\label{powmnp}
\Phi_{lm} &= (2 (m_{\rm eff})_{lm,\,lm})^2, 
\\
\label{powmnt}
(g_{\rm eff})^{tt}_{lm} &=-1/(g_{\rm eff})^{rr}_{lm} = -\frac{2(m_{\rm eff})_{kn,\,lm}}{r-2(m_{\rm eff})_{kn,\,lm}}+O(\varepsilon ^3), 
\\
\label{powmnm}
(m_{\rm eff})_{kn,\,lm} &= m+\frac{15 m}{8 \pi r}{\cal I}^{C}_{kn,\,lm}\varepsilon ^2 +O(\varepsilon ^3).
\end{align}
\end{subequations}
(For $r$-dependence in $(m_{\rm eff})_{kn,\,lm}$, see (\ref{sdghc})).  
Whether the 2D near-horizon effective action can be obtained or not is non-trivial 
before trying as mentioned in Sec.\ref{sec:intro}, to check which is one of the motivations in this study.

\section{Result of Hawking flux with  supertranslation correction}  
\label{sec:2d}

We obtain Hawking flux 
by anomaly cancellation method \cite{Robinson:2005pd,Iso:2006wa}, 
in which reducing to 2D is crucial, 
because analysis is performed with the 2D anomaly. 
For details, see Sec.\ref{sec:hfac}. 
\newline

Anomaly cancellation will focus on the fact:  
{\bf 1)} At the classical level, 
there is no outgoing flux in the near-horizon region 
for the strong gravitational effect, 
{\bf  2)} however, at the quantum level, outgoing flux will arise by the quantum tunneling \cite{Parikh:1999mf}.  
Hence, the outgoing flux exists in the near-horizon region finally. 
At this time, if one takes  in the analysis as
\begin{eqnarray}\label{lsdakl}
\textrm{amount of flux from tunneling}=\textrm{amount of lack of flux at the classical level}, 
\end{eqnarray}
the amount of the flux by the quantum tunneling can be identified as the Hawking flux. 
\newline

The amount of the outgoing flux is represented 
by the integral constant obtained from the formulas of the 2D  anomaly: 
\begin{eqnarray} \label{lfybks}  
\nabla_\mu T^\mu{}_{\nu,\, lm} = \,
-\frac{\partial_\nu \Phi_{lm}}{\sqrt{-(g_{\rm eff})_{lm}}}\frac{\delta S_{\textrm{2D}}}{\delta_L \Phi_{lm}}
+ \textrm{both/either $\mathscr{A}_{\nu,\, lm}^{\pm}$}, 
\end{eqnarray} 
which can be fixed by the condition that 
the system is symmetric,  
which is at the point where the variation of the action vanishes: 
\begin{eqnarray}\label{nvelve}
(\delta S_{\textrm{2D}})_{lm} = -\int d^2x \sqrt{-(g_{\rm eff})_{lm}}\,\eta^\nu \nabla_{\mu,\,lm} T^\mu{}_{\nu,\,lm}. 
\end{eqnarray}

The Hawking flux we have obtained has been (\ref{ea0ahmn}), $\pi T_H^2/12$,  
which is the same with just a Schwarzschild. 
The reason of this is written in Sec.\ref{subsec:femt}.  

\section{Conclusion}
\label{klsdlsd}

Although the position of the horizon has been displaced  
and whether the near-horizon field theories can reduce to free 2D  has been non-trivial, 
Hawking temperature and flux have been obtained without changes. 
This no changes in  Hawking temperature had been already clear 
when near-horizon metrices are obtained,  
however whether Hawking flux can be obtained without any changes or not is unclear for the reason written in Sec.\ref{sec:intro}.

Furthermore, although there are works relating with this study as mentioned in Sec.\ref{sec:intro},  
for the research situation mentioned there, the analysis in this study would be worthwhile.
\newline

The value of the Hawking flux would be always (\ref{ea0ahmn}) as long as the function $C$ is 
{\bf 1)} to the second-order, and 
{\bf 2)} independent of $\phi$. 
The reason for this is as follows.  

First, if the correction is to $\varepsilon^2$ 
but $C$ is some one other than (\ref{cdsvds}) independent of $\phi$, 
\begin{itemize}
\item  
highly complicated terms of $\theta$ will be newly involved into the each coefficient of $\varepsilon^{1,2}$ in (\ref{vnksld}). 
At this time, the feasibility of the integrate out for ($\theta$, $\phi$) is the problem, 
however it would be no problem by using the following formula and (\ref{fmry3}):  
\begin{equation}\label{jksafd}
Y^{m_1}_{l_1}
Y^{m_2}_{l_2}
=
\sum_{L,M}
\sqrt{\frac{(2l_1+1)(2l_2+1)}{4\pi(2L+1)}} 
\langle l_1\,0 \, l_2\,0| L\,0 \rangle
\langle l_1m_1 \, l_2 m_2 | LM \rangle Y^M_L,
\end{equation}\vspace{-5.0mm}
\item
coefficients of $\varepsilon^{1,2}$ in (\ref{powmnm}) will get highly involved concerning $\theta$,     
however the structure of (\ref{powmnt}) as the function $f$ would be no changed (see the last of Sec.\ref{subsec:femt}), 
since $C$ depends only on $\theta$ and $\phi$ by definition. 
\end{itemize}
Therefore, 
if the two conditions above are satisfied, 
one could always get the 2D near-horizon effective action 
with the $f$ same  with (\ref{bqws}) as the structure.  
On the other hand, 
\begin{itemize} 
\item
if the correction of $\varepsilon$ were involved more than 3rd-order, 
the feasibility of the analysis to get the free 2D theory as in Sec.\ref{subsec:emnh} gets unclear. 
See the last line in Sec.\ref{subsec:emnh}.  
Namely, if the same behavior with (\ref{rrwjb}) were not held, 
the analysis to get the free 2D theory would be impossible. 

\item
If $\phi$-dependence were mixed in the $C$, 
the formulas (\ref{jksafd}) might get unavailable, 
and we could not get the 2D action like (\ref{bqws}). 
\end{itemize}

This study has considered only $Y_2^0$ for the NG boson field of the supertranslation, 
then given a conclusion considering the contribution of other modes would be qualitatively same. 
Therefore this study  should be careful on whether other mode's contributions are qualitatively the same or not, 
which point is cared as above.

It is considered from our result, no changes, that 
Hawking temperature and flux may be the conserved quantities under the asymptotic symmetry.

\vspace{-3mm}
\section*{Acknowledgment} 
\vspace{-1mm}
Upon writing this article, 
many discussions with Prof.Feng-Li Lin, the collaborator in \cite{Lin:2020gva}, were very helpful. 
Using this opportunity, I would like to offer my thanks to Prof.Hai-Qing Zhang 
who could financially support the conference. 

\newpage

\noindent
{\bf \Large Appendix} 
\vspace{4mm}
\appendix 

\vspace{-8.5mm}  

\section{Metrices with supertranslation correction } 
\label{sec:mgwc} 
 
This appendix is the detailed description of \cite{Lin:2020gva}, and in this section, 
we obtain  the metrices for a 4D Schwarzschild black hole spacetime 
with supertranslations to the second-order.  

\subsection{Introduction of supertranslation} 
\label{subsec:ivst}  
 
We start with the following coordinate system for a 4D Schwarzschild black hole spacetime: 
\begin{eqnarray}\label{kcnh} 
ds^2=-(1-{2m}/{r_s})dt_s^2+(1-{2m}/{r_s})^{-1}dr_s^2+r_s^2d\Omega_s^2.    
\end{eqnarray}
We refer to this type of coordinate system as the ``Schwarzschild coordinates''. 
In order to involve the supertranslations,    
we rewrite (\ref{kcnh}) into (\ref{vpoe}),  
where $r_s\!=\!\rho_s\left(1+{m}/{2\rho_s}\right)^2$. 
We refer to this type of expression as ``isotropic coordinates''.  

\begin{figure}[htbp!!!!!]
\begin{center}  
\includegraphics[width=41mm]{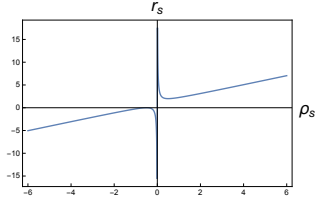} 
\end{center}\vspace{-5mm}
\caption{Plot for $r_s=\rho_s\left(1+{m}/{2\rho_s}\right)^2$ for $m=1$.}
\label{fig:one}
\end{figure}

Note that in this relation, two $\rho_s$ correspond to one $r_s$ as 
\begin{eqnarray}\label{slrh} 
\rho_s= \big( -m+r_s\pm \sqrt{-2mr_s+r_s^2} \big)/2. 
\end{eqnarray}   
See Fig.\ref{fig:one}.   
We can see   
{\bf 1)} positions of horizon in isotropic and Schwarzschild coordinates correspond each other,  
{\bf 2)} isotropic coordinates do not cover the inside of the horizon. 

We denote the supertranslated isotropic coordinates as $(t,\rho, \theta,\phi)$. 
These and $(t_s,x_s, y_s,z_s)$ are related like (\ref{ctki0}), where
\begin{eqnarray}\label{rltrtxyz}
\textrm{$d\rho_s^2+\rho_s^2 d\Omega_s^2$ in (\ref{vpoe})} = dx_s^2 + dy_s^2 + dz_s^2, \quad \textrm{$t_s$ in (\ref{vpoe})} =t.
\end{eqnarray}
We take $C$ we consider as (\ref{cdsvds}).
Description here overlaps with the one under (\ref{cdsvds}) 
(but one comment; 
if we employ $Y_1^0$ as $C$,  
r.h.s. of (\ref{rltrtxyz}) results in just $d\rho^2+\rho^2d\theta^2+\rho^2\sin^2 \theta d\phi^2$,
namely no supertranslation corrections, which get involved from (\ref{ctki21}), supertranslated $\rho$.)

\subsection{Isotropic coordinates with supertranslations} 
\label{subsec:iswst} 

We now write (\ref{vpoe}) in terms of $(t,\rho, \theta,\phi)$. 
For the parts in (\ref{vpoe}), we can write as
\begin{subequations} 
\begin{align}
\label{ctki21}
\rho_s^2 
=& \,\,
x_s^2+y_s^2+z_s^2,\\*
\label{ctki22}
( 1 + {m}/{2\rho_s} )^4 ( d \rho_s^2 + \rho_s^2 d \Omega_s^2 )  
=& \,\,
g_{\rho\rho}   d\rho^2 + g_{\theta\theta}d\theta^2  + g_{\phi\phi}d\phi^2.
\end{align}
\end{subequations}
We can evaluate $d \rho_s^2 + d\rho_s^2 d \Omega_s^2$ as (\ref{rltrtxyz}), 
with which we obtain $g_{MN}$ from now. 
We write 
\begin{subequations}\label{ksdvd} 
\begin{align}
&\textrm{$g_{MN}$ for the metrices of the supertranslated isotropic coordinates} \\
&\textrm{$j_{MN}$ for the metrices of the supertranslated Schwarzschild coordinates} 
\end{align}
\end{subequations}  
in what follows, where $M,N$ in $g_{MN}$ and $j_{MN}$ refer to $(t,\rho, \theta,\phi)$ and $(t,r, \theta,\phi)$.

We can obtain $\rho_s$ by calculating (\ref{ctki21}) using (\ref{ctki0}) to $\varepsilon^2$-order as
\begin{eqnarray} \label{ctki21}
\rho_s
= 
\rho
-\frac{1}{8} \sqrt{\frac{5}{\pi }} \varepsilon m (3 \cos (2 \theta )+1)
+\frac{45 \varepsilon ^2 m^2 \sin ^2(2 \theta )}{32 \pi  \rho }
+O(\varepsilon ^3).
\end{eqnarray}  
With this we can  obtain $g_{MN} $ as 
\begin{subequations} 
\label{ctkixx}   
\begin{align} 
\label{ctkitt}  
\bullet \hspace{2mm}
g_{tt} 
\! =& \, 
-\frac{(m-2 \rho )^2}{(m+2 \rho )^2}
-\sqrt{\frac{5}{\pi }} \varepsilon  m^2\frac{  (m-2 \rho )}{(m+2 \rho )^3}(3 \cos (2 \theta )+1) 
-\frac{5 \varepsilon ^2 m^3 }{8 \pi  \rho  (m+2 \rho )^4}
(22 m \rho -9 m^2 
\nonumber\\*
& \,\,
+14 \rho ^2
+9 \cos (4 \theta ) (m^2+2 m \rho -6 \rho ^2)
+24 \rho  \cos (2 \theta ) (m-\rho )
)
+O(\varepsilon ^3),
\\   
\label{ctkirr}  
\bullet \hspace{2mm}
g_{\rho\rho} 
\! =& \, 
\frac{(m+2 \rho )^4}{16 \rho ^4}
+\sqrt{\frac{5}{\pi }} \varepsilon  m^2\frac{ (m+2 \rho )^3}{32 \rho ^5} (3 \cos (2 \theta )+1)
+\frac{5 \varepsilon ^2 m^3 (m+2 \rho )^2}{1024 \pi  \rho ^6}(19 m-28 \rho
\nonumber\\*
& \,\,
+12 \cos (2 \theta ) (5 m+4 \rho )+27 \cos (4 \theta ) (3 m+4 \rho )
)
+O(\varepsilon ^3),
\\
\label{ctkithth}   
\bullet \hspace{2mm}
g_{\theta\theta}
\! =& \, 
\frac{(m+2 \rho )^4}{16 \rho ^2}
+ 
\sqrt{\frac{5}{\pi }} \varepsilon m (m+2 \rho )^3  \, 
\frac{  3 \cos (2 \theta ) (5 m+6 \rho )+m-2 \rho }{64 \rho ^3}
+\frac{5 \varepsilon ^2 m^2 (m+2 \rho )^2}{2048 \pi  \rho ^4}(
\nonumber\\*
& \hspace{4mm}
12 \cos (2 \theta ) (39 m^2+76 m \rho +20 \rho ^2)+9 \cos (4 \theta ) (27 m^2+44 m \rho +4 \rho ^2)
\nonumber\\*
&  \,\,
+249 m^2+484 m \rho +236 \rho ^2
)
+O(\varepsilon ^3),
\\
\label{ctkipp} 
\bullet \hspace{2mm}
g_{\phi\phi}
\! =& \, 
\frac{ (m+2 \rho )^4\sin ^2(\theta )}{16 \rho ^2}
+\sqrt{\frac{5}{\pi }} \varepsilon  m \frac{ (m+2 \rho )^3}{64 \rho ^3}  \sin ^2(\theta )
(
\cos (2 \theta ) (9 m+6 \rho )+7 m+10 \rho
)
\nonumber\\*
& \,\,
+\frac{5 \varepsilon ^2 m^2 \sin ^2(\theta ) (m+2 \rho )^2}{2048 \pi  \rho ^4}
(
249 m^2+484 m \rho +236 \rho ^2 
+12 \cos (2 \theta ) (39 m^2+76 m \rho 
\nonumber\\*
& \,\,
+20 \rho ^2)
+9 \cos (4 \theta ) (27 m^2+44 m \rho +4 \rho ^2)
)
+O(\varepsilon ^3).
\end{align}
\end{subequations}

\subsection{Rewriting from Schwarzschild to isotropic coordinates} 
\label{subsec:rwscisc} 

Since we have obtained the metrices in the supertranslated isometric coordinates $(t,\rho,\theta,\phi)$,  
we will rewrite these to the following Schwarzschild coordinates:  
\begin{eqnarray}\label{schcdst}
ds^2
=
-( 1-{2 \mu }/{r} ) dt^2  
+(1-{2 \mu }/{r})^{-1}dr^2
+ j_{\theta\theta} d\phi^2
+ j_{\phi\phi} d\phi^2.
\end{eqnarray} 
Then, we will find the mass part $\mu$ cannot remain constant 
(if analysis is to $\varepsilon^1$-order, it can be constant). 
Therefore, we treat $\mu$ as $\mu(\rho)$. 
In what follows, we obtain 
{\bf 1)} a relation between $r$ and $\rho$, and 
{\bf 2)} $\mu(\rho)$ as the solution, 
by solving the following relations:
\begin{subequations}
\begin{align}
\label{ccisa1}
\bullet & \quad  \!\! - ( 1-{2 \mu(\rho) }/{r} ) =\, g_{tt},\\*
\label{ccisa2} 
\bullet & \quad \!\! \frac{1}{1-{2 \mu(\rho) }/{r}} \Big( \frac{dr}{d\rho} \Big)^2 =\, g_{\rho\rho}.
\end{align}
\end{subequations}
The argument in $\mu(\rho)$ should be $\rho$. 
If we express $\mu(\rho)$ in terms of $r$, (\ref{ctfsti1}) is plugged in.

We can obtain the $r$ satisfying (\ref{ccisa1}) to $\varepsilon^2$-order as
\begin{eqnarray}\label{ctfits} 
\!\!
r
\!\!\! &=& \!\!\!
\frac{\mu (\rho ) (m+2 \rho )^2}{4 m \rho } 
+\frac{\varepsilon  }{32 \rho ^2}\sqrt{\frac{5}{\pi }} (3 \cos (2 \theta )+1) \mu (\rho ) (m^2-4 \rho ^2)
\nonumber\\*
\!\!\! && \!\!\! +
\frac{5 \varepsilon ^2 m \mu (\rho )}{512 \pi  \rho ^3}
(9 \cos (4 \theta ) (3 m^2-8 \rho ^2)+12 m^2 \cos (2 \theta )-7 m^2+72 \rho^2)
+O(\varepsilon ^3).  
\end{eqnarray}

Let us obtain the $\mu(\rho)$. 
For this, 
look (\ref{ccisa2}), 
then plugging (\ref{ctfits}) into the $r$, 
solve it for $\mu(\rho)$ order by order to $\varepsilon^2$-order. 
As a result we can obtain 
\begin{eqnarray}
\label{micfit1} 
\mu(\rho) 
\!\!\! &=& \!\!\!
m
+\frac{c_1 \rho \varepsilon}{(m+2 \rho )^2}
+
\frac{\varepsilon ^2}{64 \pi  \rho ^2 (m+2 \rho )^3}
\{
-24 \sqrt{5 \pi } c_1 m \rho ^2 \cos (2 \theta ) (m-2 \rho )
\nonumber\\*
&& \!\!\!
+ \, 8 \rho ^2 (8 \pi  c_2 \rho  (m+2 \rho )
-\sqrt{5 \pi } c_1 m (m-2 \rho ))
\nonumber\\*
&& \!\!\!
+ \, 30 m^3 \sin^2(2 \theta ) (m+2 \rho ) (m^2-12 \rho ^2)
\}
+O(\varepsilon ^3),
\end{eqnarray}  
where we took the integral constant at $\varepsilon^0$-order so that $\varepsilon^0$-order becomes $m$. 
$c_{1,2}$ are the integral constants at $\varepsilon^{1,2}$-orders 
(these can depend on ($\theta$, $\phi$) and determined at (\ref{dtitc})).

Now we have obtained the relation ``$r=\cdots$'' as in (\ref{ctfits}), 
with which, 
rewriting the Schwarzschild to the isotropic coordinates to $\varepsilon^2$-order is possible:
\begin{eqnarray}\label{rwfsti}
&& \hspace{-8.5mm}
-(1-{2\mu(\rho)}/{r})dt^2
+(1-{2\mu(\rho)}/{r} )^{-1}dr^2
+j_{\theta\theta}d\theta^2 + j_{\phi\phi}d\phi^2 \nonumber\\*
&& \hspace{-12mm}
\rightarrow
g_{tt}dt^2
+g_{\rho\rho}d\rho^2
+\Big( 
j_{\theta\theta}+\frac{1}{1-\frac{2\mu(\rho)}{r}}\Big(\frac{\partial r}{\partial \theta}\Big)^2
\Big)d\theta^2
+\frac{2}{1-\frac{2\mu(\rho)}{r}}\frac{\partial r}{\partial \rho}\frac{\partial r}{\partial \theta}d\rho d\theta
+j_{\phi\phi}d\phi^2\,.
\end{eqnarray}
However what is needed is rewriting from the isotropic to Schwarzschild coordinates.

\subsection{Rewriting from isotropic to Schwarzschild coordinates} 
\label{subsec:reisscco} 

We will obtain the relation between $\rho$ and $r$ in the form ``$\rho = \cdots$''  
to $\varepsilon^2$-order 
to become possible to rewrite  (\ref{rwfsti}) in the opposite direction.    
For this, there are two ways: to solve
{\bf 1)} (\ref{ccisa2}) or {\bf 2)} (\ref{ctfits}). 
As a result of our try, if we solve to $\varepsilon^1$-order, 
we can get the same $\rho$ from either of them (we checked this sameness numerically). 
However, if we try to obtain to $\varepsilon^2$-order, we can obtain only from 2) 
(for some technical reason of mathematica). 

Writing what we did, 
plugging $\mu(\rho)$ in (\ref{micfit1}) into the $\mu$ in (\ref{ctfits}), 
then expanding it to $\varepsilon^2$-order, 
we can obtain $\rho$ order by order. 
As a result, four solutions are obtained. 
At this time, the $\varepsilon^0$-order in the two of these do not agree with (\ref{slrh}), 
while those of the rest two can agree with (\ref{slrh}). 
Therefore, we employ the latter two, which are 
\begin{eqnarray}\label{ctfsti1}
\rho^{(1,2)}(r)
\!\!\!\! &=& \!\!\!\!
\frac{1}{2} (r-m \mp\sqrt{r (r-2 m)} )
+\frac{\varepsilon }{8 \pi  m r (2 m-r)}
\{
\pi  c_1 r (r-2 m)
\pm \pi  c_1 m \sqrt{r (r-2 m)}
\nonumber\\*
\!\!\!\! && \!\!\!\!
\mp\pi c_1 r \sqrt{r (r-2 m)}
-\sqrt{5 \pi } m^2 r (3 \cos (2 \theta )+1) (r-2 m)
\}
-\frac{\varepsilon ^2}{128 \pi  m r^2 (r-2 m)^2}
[
\nonumber\\*
\!\!\!\! && \!\!\!\!
16 \pi  c_2 (r^4\mp \sqrt{r^7 (r-2 m)})
-120 m^4 r (4 r\pm \sqrt{r (r-2 m)})
\nonumber\\*
\!\!\!\! && \!\!\!\!
+60 m^3 (16 r^3\pm 9 \sqrt{r^5 (r-2 m)})
-8 m^2 \{(60 \sqrt{r^7 (r-2 m)}\mp 8 \pi  c_2 r^2)
+r^4 \pm 75
\nonumber\\*
\!\!\!\! && \!\!\!\!
+4 \pi  c_2 r \sqrt{r (r-2 m)}\}
+m \{120 r^5\pm 120 \sqrt{r^9 (r-2 m)}\mp 2 \pi  c_1{}^2 \sqrt{r (r-2 m)}
\nonumber\\*
\!\!\!\! && \!\!\!\!
-64 \pi  c_2 r^3 
\pm 48 \pi  c_2 \sqrt{r^5 (r-2m)}\}
+60 m r (2 m-r) \cos (4 \theta )  \{
\nonumber\\*
\!\!\!\! && \!\!\!\!
m^2 (4 r\pm \sqrt{r (r-2 m)})
+2 (r^3\pm \sqrt{r^5 (r-2 m)})
-2 m r (3 r\pm 2 \sqrt{r (r-2 m)})
\}
\nonumber\\*
\!\!\!\! && \!\!\!\!
]
+O(\varepsilon ^3),
\end{eqnarray}
where the $1$ and $2$ in the $\rho^{(1,2)}(r)$ correspond as
\begin{eqnarray}\label{nmcrdt}
\textrm{(1,2) $\to$ $(+,-)$ of $\pm$ and $(-,+)$ of $\mp$.} 
\end{eqnarray}

Let us determine which $\rho^{(1,2)}(r)$ we employ and determine $c_{1,2}$. 
For this, plugging $\rho^{(1,2)}$ in (\ref{ctfsti1}) into 
$\mu(\rho)$ in (\ref{micfit1}),  
write it in terms of $r$ to $\varepsilon^2$-order as 
\begin{eqnarray}
\label{ctfstu1}
\mu^{(1,2)}\big(\rho^{(1,2)}(r)\big)
\!\!\!\! &=& \!\!\!\!
m
+\frac{c_1 \varepsilon }{4 r}
+\frac{\varepsilon^2}{8 \pi  m r (\sqrt{r (r-2 m)}\mp r)^3 (\sqrt{r (r-2 m)}\pm m\mp r)^2}
[
\nonumber\\*
&& \!\!\!\! 
\pi   
\{
-4 c_2 m^4 r \sqrt{r (r-2 m)}
+48 c_2 m^3 \sqrt{r^5 (r-2 m)}
-c_1{}^2 m^3 \sqrt{r (r-2 m)}
\nonumber\\*
&& \!\!\!\!
-80 c_2 m^2 \sqrt{r^7 (r-2m)}
+12 c_1{}^2 m^2 r \sqrt{r (r-2 m)}
+32 c_2 m \sqrt{r^9 (r-2 m)}
\nonumber\\*
&& \!\!\!\!
+r (\pm 7 m^3 \mp 28 m^2 r \pm 28 m r^2 \mp 8 r^3) (4 c_2 m r+c_1{}^2)
+8 c_1{}^2 \sqrt{r^7 (r-2 m)}
\nonumber\\*
&& \!\!\!\!
-20 c_1{}^2m \sqrt{r^5 (r-2 m)}
\}
-30 m^4 r \sin^2 (2 \theta ) 
\{
m^2 (\sqrt{r (r-2 m)}\mp7 r)
\nonumber\\*
&& \!\!\!\!
\pm 3mr (5r \mp 3 \sqrt{r (r-2 m)})
+6( \mp r^3 + \sqrt{r^5 (r-2 m)})
\}
]
+O(\varepsilon ^3).
\end{eqnarray}
Behavior of these at the distant region is given as 
\begin{subequations}
\begin{align}
\label{ctfexu1}
\mu^{(1)}\big(\rho^{(1)}(r)\big)
\! =& \,
m
+\varepsilon  
\big(
\frac{c_1}{4 r}+O\left(r^{-2}\right)
\big)
+\varepsilon ^2 
\Big(
\frac{15  r^2\sin ^2(2\theta )}{2\pi m  }
-\frac{45 \sin ^2(2\theta ) r}{2\pi }
+\frac{45 m \sin ^2(2\theta )}{4\pi }
\nonumber\\*
&
+\frac{15 m^2 \sin ^2(2 \theta )/\pi +c_2}{4r}
+O\left(r^{-2}\right)
\Big)
+O(\varepsilon ^3),
\\
\label{ctfexu2}
\mu^{(2)}\big(\rho^{(2)}(r)\big)
\! =& \,
m
+\varepsilon 
\big(
\frac{c_1}{4 r}+O\left(r^{-6}\right)
\big)
+ 
\frac{\varepsilon ^2}{4} 
\Big( 
\frac{c_2}{r}
+\frac{2 c_1{}^2-45 m^4 \sin ^2(2 \theta )/\pi }{8 mr^2}
-\frac{45 m^4 \sin ^2(2 \theta )}{8 \pi  r^3} 
\nonumber\\*
&
-\frac{105 m^5 \sin ^2(2 \theta )}{16 \pi r^4}
-\frac{135 m^6 \sin ^2(2 \theta )}{16 \pi r^5}
+O\left(r^{-6}\right)
\Big)
+O\left( \varepsilon ^3\right).
\end{align}
\end{subequations}
It can be seen from the above we should discard $\mu^{(1)}$ by the reason: 
It is always diverged at the distant region irrelevantly of $c_{1,2}$
for the terms, $\frac{15 r^2 \sin ^2(2\theta )}{2\pi  m}-\frac{45 r \sin ^2(2\theta ) }{2\pi }$.
Thus, it is enough only with  $\mu^{(2)}$ in the following, but we proceed with both  just in case.  
\newline

Now we determine $c_{1,2}$. 
Since these are integral constants, we are allowed to take these arbitrarily. 
However in this study, by the reason written in what follows we will  take as
\begin{eqnarray}\label{dtitc}
c_{1,2}=0. 
\end{eqnarray}

Looking $\mu^{(2)}(\rho^{(2)}(r))$, 
we can find that 
it diverges at $m=0$ unless $c_1$ is zero
for the term $\frac{2c_1{}^2}{32 m r^2}$ at its $\varepsilon^2$-order. 
Hence we take $c_1$ to $0$.

As for our logic for $c_2$,  
{\bf 1)}   
consider starting with just a flat spacetime patched  
by \linebreak Schwarzschild coordinates with the zero mass, therefore $\mu^{(2)}$ at the starting stage is zero.  
{\bf 2)}  
Suppose changing it to the isotropic coordinates,  
involve the supertranslations. 
Then, back the coordinates to the Schwarzschild.  
At this time, the expanded $\mu^{(2)}$ is given by (\ref{ctfexu2}). 
{\bf 3)}  
At this time, the mass should be zero, 
therefore $\mu^{(2)}$ should be zero.  
However, if $c_2$ is not zero, 
we can see $\mu^{(2)}$ is not zero for the terms $\frac{c_1}{4 r}$ at the $\varepsilon^1$-order 
and $\frac{c_2}{4 r}$ at the $\varepsilon^2$-order. 
{\bf 4)} 
As $c_1$ has been taken to zero in the above, 
we take $c_2$ to zero. 

Above, we have considered 
in terms of the supertranslation toward the flat spacetime 
and based on the consideration that 
mass in the spacetime should not be changed by the supertranslation. 
The same issue is taken up in Sec.24.2 in \cite{Lust:2018cvp}.  
There, again mass is not introduced, 
though $C_{zz}$ and $C_{\bar{z}\bar{z}}$ are introduced.

With (\ref{dtitc}), 
(\ref{ctfsti1}) and (\ref{ctfstu1}) are fixed as
\begin{eqnarray}
\label{trrlcx1}
&& \!\!\!\!\!\!\!\!\!\!\!\!\!\!\!\!\!\!
\rho^{(1,2)} 
=
\frac{1}{2} 
( \mp \sqrt{r (r-2 m)}-m+r)
+\frac{1}{8} \sqrt{\frac{5}{\pi }} \varepsilon  m (3 \cos (2 \theta )+1)
-\frac{15 \varepsilon ^2 \sin^2 (2 \theta ) }{16 \pi  r^2 (r-2m)^2}
\{
\nonumber\\*
&& \hspace{1mm}
-10 m r^4
\mp 8 m \sqrt{r^7 (r-2 m)}
+2(r^5 \pm  \sqrt{r^9 (r-2 m)})
-2 m^3 r(4 r\pm \sqrt{r (r-2 m)})
\nonumber\\*
&& \hspace{1mm}
+m^2 (16 r^3 \pm 9 \sqrt{r^5 (r-2 m)})
\}
+O(\varepsilon ^3),
\\
\label{ctfucx1}
&&  \!\!\!\!\!\!\!\!\!\!\!\!\!\!\!\!\!\!
\mu^{(1,2)}(\rho^{(1,2)}(r))
=
m
+
\frac{
15 \varepsilon ^2 m^3 \sin ^2(2 \theta ) 
( m^2-3 (\sqrt{r (r-2 m)}\pm m \mp  r)^2)
}
{8 \pi ( r \mp \sqrt{r (r-2 m)})^2 ( \sqrt{r (r-2m)}\pm m\mp r )^2 } 
+O(\varepsilon ^3).
\end{eqnarray}
Using these we can rewrite the isotropic to the Schwarzschild coordinates as
\begin{eqnarray}\label{rwfits}
\!\!\!\!\!\! && \!\!\!\!
g_{tt}dt^2
+g_{\rho\rho}d\rho^2
+g_{\theta\theta}d\theta^2
+g_{\phi\phi}d\phi^2 
\nonumber\\*  
\!\!\!\!\!\! &\rightarrow&  \!\!\!\!
-\Big(1-\frac{2\mu(\rho)}{r} \Big) dt^2
+\Big(1-\frac{2\mu(\rho)}{r} \Big)^{-1} dr^2
+\Big( g_{\theta\theta}+g_{\rho\rho}\Big(\frac{\partial \rho}{\partial \theta}\Big)^2 \Big)d\theta^2
\nonumber\\*
&& \!\!\!\! + \,   
2g_{\rho\rho}\frac{\partial \rho}{\partial r}\frac{\partial \rho}{\partial \theta}d\rho d\theta
+ g_{\phi\phi}d\phi^2 
\nonumber\\*
&& \!\!\!\! 
\equiv
j_{tt} dt^2 
+j_{rr} dr^2
+ j_{\theta\theta} d\theta^2 
+ 2j_{r\theta} dr d\theta
+j_{\phi\phi}d\phi^2,
\end{eqnarray}
where $\rho=\rho(r,\theta)$ and $g_{MN}$ are  in (\ref{ctkixx}). 
We give the expressions of $j_{MN}$ in the next. 

\subsection{Metrices with correction to $\varepsilon^2$-order in 4D Schwarzschild coord.} 
\label{subsec:ormt} 
 
We give the expression of $j_{MN}$ in (\ref{rwfits}) in the case of  (\ref{dtitc}). 
\begin{subequations}
\begin{align}
\hspace{-4.0mm}
\bullet \hspace{2mm}
\label{ctks001}  
j^{(1,2)}_{tt} 
\! =& \,
-( 1-\frac{2 m}{r} )
+
\frac{
15 \varepsilon ^2 \sin ^2(2 \theta ) 
(m^5-3 m^3 (\sqrt{r (r-2 m)}\pm m \mp r)^2)
}{
4 \pi  r (r \mp \sqrt{r (r-2 m)})^2 
( \sqrt{r (r-2m)}\pm m\mp r)^2
}
+O(\varepsilon ^3),
\end{align}
\end{subequations}
\vspace{-6.0mm}
\begin{subequations}
\begin{align}
\hspace{-12.0mm}
\bullet \hspace{2mm}
\label{ctks111}
j^{(1,2)}_{rr} 
\! =& \,
(1-\frac{2 m}{r} )^{-1}
+
\frac{15 \varepsilon ^2 \sin ^2(2 \theta ) }
{4 \pi  m (r-2m)^2}
\{
m^3+3 m^2 r-6 m r^2
\pm2 (\sqrt{r^5 (r-2 m)}\pm r^3)
\nonumber\\*
&
\mp 4 m r \sqrt{r (r-2 m)}
\}
+O(\varepsilon ^3),
\end{align} 
\end{subequations}
\vspace{-6.0mm}
\begin{subequations}
\begin{align}
\bullet \hspace{2mm}
\label{ctks221} 
j^{(1)}_{\theta\theta} 
\! =& \,
r^2
- \frac
{3 \sqrt{\frac{5}{\pi }} \varepsilon  m \cos (2 \theta )(r - \sqrt{r (r-2 m)})^4}
{2 (\sqrt{r (r-2 m)} + m - r)^3}
+\frac{15 \varepsilon ^2 m^2}{8 \pi  r (r-2 m)^2 (\sqrt{r (r-2 m)}+m-r)^3}
\nonumber\\*
&
\times
\{r^2 (r-2 m)^2(\cos (4\theta )(m^2+6 m r-9 r^2)-(m^2-18 m r+15 r^2))
\nonumber\\*
&
+12 m^3 \sqrt{r^5 (r-2 m)} (\cos (4 \theta ) -1)
+ m^2 \sqrt{r^7 (r-2 m)} (24\cos (4 \theta ) +72)
\nonumber\\*
&
- m \sqrt{r^9 (r-2 m)} (33\cos (4 \theta )+63 ) 
+ \sqrt{r^{11} (r-2 m)} (9\cos (4 \theta )+15 )
\}
+O(\varepsilon ^3),
\\[-2.5mm]
\label{ctks222} 
j^{(2)}_{\theta\theta} 
\! =& \,
r^2
+\frac{3 \sqrt{\frac{5}{\pi }} \varepsilon  m \cos (2 \theta ) (\sqrt{r (r-2 m)}+r)^4}{2 (\sqrt{r (r-2 m)}-m+r)^3}
\nonumber\\*
&
+\frac{}
{
8 \pi  r (r-2 m)^2 
\{ 
m^4-4 m^3 (\sqrt{r (r-2 m)}+4 r)+20 m^2 r (\sqrt{r (r-2 m)}+2 r) 
}
\nonumber\\*
&
\frac{15 \varepsilon ^2 m^2}
{
-24 m \sqrt{r^5 (r-2 m)}-32 m r^3
+8 (\sqrt{r^7(r-2 m)}+r^4)\}
}
[
\nonumber\\*
&
-4 m^5 r^2+8 m^4 
(2 \sqrt{r^5 (r-2 m)}+13 r^3)
-5 m^3 (32 \sqrt{r^7 (r-2 m)}+73 r^4)
\nonumber\\*
&
+m^2 (268 \sqrt{r^9 (r-2 m)}+409r^5)+\cos (4 \theta ) 
\{
4 m^5 r^2-8 m^4 (2 \sqrt{r^5 (r-2 m)}+r^3)
\nonumber\\*
&
+m^3 (-32 \sqrt{r^7 (r-2 m)}-115 r^4 )
+m^2 (116 \sqrt{r^9 (r-2 m)}+191r^5 )
\nonumber\\*
&
-84 m \sqrt{r^{11} (r-2 m)}-102 m r^6+18 (\sqrt{r^{13} (r-2 m)}+r^7)
\}
\nonumber\\*
&
+30 (\sqrt{r^{13} (r-2 m)}+r^7)
-6 m (26 \sqrt{r^{11} (r-2 m)}+31r^6)
]
+O(\varepsilon ^3), 
\end{align} 
\end{subequations}
\vspace{-6.0mm}
\begin{subequations}
\begin{align}
\bullet \hspace{2mm} 
\label{ctks121}
j^{(1,2)}_{r\theta} 
\! =& \,
-\frac{3 \sqrt{\frac{5}{\pi }} \varepsilon  m \sin (2 \theta ) 
(
r\mp\sqrt{r (r-2m)}
)^4}
{8 \sqrt{r (r-2 m)} (\sqrt{r (r-2 m)} \pm m \mp r )^3}
\pm\frac{15 \varepsilon ^2 m^4 r \sin (4 \theta )}{4 \pi  (2 m-r) (\sqrt{r (r-2 m)} \pm m \mp r )^3}
\nonumber\\*
&
+O(\varepsilon ^3),
\end{align}
\end{subequations}
\vspace{-6.0mm}
\begin{subequations} 
\begin{align}
\bullet \hspace{2mm} 
\label{ctks331} 
j^{(1)}_{\phi\phi} 
\! =& \,
r^2 \sin ^2(\theta )
-\frac{3 \sqrt{\frac{5}{\pi }} \varepsilon  m \sin ^2(2 \theta ) (r-\sqrt{r (r-2m)})^4}{8 (\sqrt{r (r-2 m)}+m-r)^3}\nonumber\\*
&+\frac{15 \varepsilon ^2 m^2 \sin ^2(2 \theta ) }{8 \pi  r (r-2 m)^2 (\sqrt{r (r-2 m)}+m-r)^3}
\{
12 m^3 \cos (2 \theta ) \sqrt{r^5 (r-2 m)}\nonumber\\*
&
-12 m^3 \sqrt{r^5 (r-2 m)}+24 m^2 \sqrt{r^7 (r-2 m)}+r^2 \cos (2 \theta ) (r-2 m)^2(m^2-3 r^2)\nonumber\\*
&
-r^2 (r-2 m)^2 (m^2-6 m r+3 r^2)+3 \cos (2 \theta ) \sqrt{r^{11} (r-2 m)}+3 \sqrt{r^{11} (r-2 m)}\nonumber\\*
&
-9 m \cos (2 \theta ) \sqrt{r^9 (r-2 m)}-15 m
\sqrt{r^9 (r-2 m)}
\}
+O(\varepsilon ^3),
\\ 
\label{ctks332} 
j^{(2)}_{\phi\phi} 
\! =& \,
r^2 \sin ^2(\theta)
+\frac{3 \sqrt{\frac{5}{\pi }} \varepsilon  m \sin ^2(2 \theta )(\sqrt{r (r-2 m)}+r)^4}{8 (\sqrt{r (r-2 m)}-m+r)^3}\nonumber\\*
&
+\frac{15 \varepsilon ^2 m^2 \sin ^2(2\theta ) }{4 \pi  r^2 (r-2 m)^2 (\sqrt{r (r-2 m)}-m+r)^4}
[
-4 m^5 r^3+8 m^4 (2 \sqrt{r^7 (r-2 m)}+7 r^4)\nonumber\\*
&
-m^3 (64 \sqrt{r^9 (r-2 m)}+125 r^5)+m^2 (76\sqrt{r^{11} (r-2 m)}+109 r^6)
\nonumber\\* 
&
+\cos (2 \theta ) 
\{
4 m^5 r^3-16 m^4 (\sqrt{r^7 (r-2 m)}+2 r^4)+m^3 (16 \sqrt{r^9 (r-2 m)}+5 r^5)
\nonumber\\*
&
+m^2 ( 20 \sqrt{r^{11} (r-2 m)}+41 r^6)+6 (\sqrt{r^{15} (r-2 m)}+r^8)
\nonumber\\*
&
-6 m (4 \sqrt{r^{13} (r-2 m)}+5 r^7)
\}
+6 (\sqrt{r^{15} (r-2 m)}+r^8)
\nonumber\\
&
-6 m(6 \sqrt{r^{13} (r-2 m)}+7 r^7)
]
+O(\varepsilon ^3).
\end{align}
\end{subequations}  
{\bf 1)} 
numbers in the superscripts mean those $j_{MN}$ are associated with which of $\rho^{(1,2)}$ in (\ref{trrlcx1}) with (\ref{nmcrdt}). 
{(}Origin of $j^{(1,2)}_{MN}$ is (\ref{ctfsti1}), 
then it turns out above (\ref{dtitc}) that $\mu^{(1)}$ 
is unphysical and $\mu^{(2)}$ is physical. 
$j^{(2)}_{MN}$ is associated with $\mu^{(2)}$.{)}  
{\bf 2)} Killing vector in the system above  is $\xi^M=(1,0,0,1)$ as well. 
With either $j^{(1,2)}_{MN}$,
{\bf 3)} Einstein eq. is satisfied to $\varepsilon^2$-order. 
Also from either $j^{(1,2)}_{MN}$, 
{\bf 4)} (\ref{ctkixx}) can be obtained using following one, (\ref{ctfits}) with (\ref{micfit1}) and (\ref{dtitc}):
\begin{eqnarray}
r \!\!\!\! &=& \!\!\!\!
\frac{m^2}{4 \rho }+m+\rho
+\frac{\sqrt{\frac{5}{\pi }}   m (m^2-4 \rho ^2)(3 \cos (2 \theta )+1) }{32 \rho ^2}\varepsilon 
-\frac{5  m^4 (1-12 \cos (2 \theta )-21 \cos (4 \theta ))}{512 \pi  \rho ^3}\varepsilon ^2
\nonumber\\
&& \!\!\!\!
+O(\varepsilon ^3).
\end{eqnarray}

\subsection{Comment on $c_{1,2}$} 
\label{vdsvsd} 

Positions of the horizon in supertranslated isotropic and Schwarzschild coordinates are 
\begin{eqnarray}
\label{lhrh2}
\rho_{h,4D} 
\!\!\! &=& \!\!\! 
\frac{m}{2}
+\frac{1}{8} \sqrt{\frac{5}{\pi }} m \varepsilon  (3 \cos (2 \theta )+1)
-\frac{45 }{16 \pi } \varepsilon ^2 m \sin ^2(2 \theta)
+O(\varepsilon ^3),
\\
\label{lhrb2}   
r_{h,4D}   
\!\!\! &=& \!\!\! 
2 m-\frac{15}{8 \pi } \varepsilon ^2 m \sin ^2(2 \theta )+O(\varepsilon ^3),
\end{eqnarray}
where $r_{h,4D}$  gets to this above regardless of $j^{(1,2)}_{MN}$. 
Then it turns out 
$\rho_{h,4D}$ can be transformed to $r_{h,4D}$ through (\ref{ctfits}) 
($\mu(\rho)$  is replaced by (\ref{micfit1})), 
but $r_{h,4D}$ is transformed to
\begin{eqnarray}\label{cddgs}
\frac{m}{2}
+\sqrt{\frac{5}{\pi }} m \varepsilon 
(\frac{3 \cos (2 \theta )+1}{8}-\frac{1}{16}\sqrt{6(-1+\cos (4\theta))})
-\frac{45 \varepsilon ^2 m \sin ^2(2 \theta)}{16 \pi } 
+O(\varepsilon ^3).
\end{eqnarray}
through (\ref{trrlcx1}).

Since $j_{MN}$ can be transformed to $g_{MN}$ by (\ref{trrlcx1}), 
$r$-coordinate corresponds to $\rho$-coordinate by (\ref{trrlcx1}).  
However, 
as mentioned above (\ref{dtitc}), there is freedom for how to take $c_{1,2}$.     
Moreover, 
as the problem of how the coordinates are patched toward a spacetime,  
there is ambiguity up to $c_{1,2}$ for the mapping of each point in $r$-coordinate to each point in $\rho$-coordinate  
through (\ref{trrlcx1}), and vice verse through (\ref{ctfits}).

Actually,  the position of horizon in the Schwarzschild coordinate is obtained  
if one proceeds calculation with unfixed $c_{1,2}$ as  
\begin{eqnarray}\label{bfeswl}
2 m  
+\frac{c_1 \varepsilon }{4m}
+\varepsilon ^2 (\frac{c_2}{4 m}-\frac{15 m \sin ^2(2 \theta )}{8 \pi })
+O(\varepsilon ^3). 
\end{eqnarray}
Here, 
the position of horizon 
in the Schwarzschild coordinate
obtained from $\rho_{h,4D}$ 
through (\ref{ctfits}) (this (\ref{ctfits}) is given with unfixed $c_{1,2}$) is (\ref{bfeswl}). 
Therefore, the position of horizon in the isotropic coordinate is always mapped 
to that in the Schwarzschild coordinate.

Toward (\ref{bfeswl}), 
if we take as
$c_1=0$ and $c_2=\frac{15m^2 \sin^2 (2 \theta )}{2 \pi }$, 
$\rho_{h,4D}$ can be obtained through  (\ref{trrlcx1}) (this (\ref{trrlcx1}) is given with these $c_{1,2}$). 
However, the $r_{h,4D}$ at that time is  $2m+O(\varepsilon ^3)$.

One may consider to determine $c_{1,2}$ 
based on agreement of the positions of horizon. 
However these should be zero for the reason under (\ref{dtitc}), 
therefore the positions of horizon determine to those obtained from $c_{1,2}=0$, (\ref{lhrh2}) and (\ref{lhrb2}).

\section{2D effective near-horizon action}
\label{sec:dra}

We have obtained the metrices 
with the supertranslation correction to the second-order 
in the Schwarzschild coordinates.  
In this section, 
obtaining the near-horizon expression of these, 
we consider the scalar field theory. 
Then, expanding the field by the spherical harmonics, 
we integrate out its ($\theta$, $\phi$). 
We will finally obtain 2D effective near-horizon action. 
The scalar field theory we consider is 
\begin{eqnarray}\label{sacf}
S_{\rm scalar} = \frac{1}{2} \int d^4x \, \sqrt{-g} \,   j^{MN}\partial_M \phi^* \partial_N \phi ,   
\end{eqnarray}
where $M,N = t, r, \theta$,  $\phi$, and $j^{MN}$ mean $j^{(2)MN}$.  
We do not include the mass and interaction terms,  
since these are ignorable 
in the near-horizon \cite{Umetsu:2010ts}\footnote{
It is considered that
theories effectively become 2D free massless 
in the near-horizon at the classical level 
as the particles effectively fall freely and these longitudinal motions get dominant.
}. 

\subsection{Near-horizon metrices}
\label{subsec:nhm}

To obtain the near-horizon expression of (\ref{sacf}), 
we first obtain the 4D metrices $j^{(2)}_{MN}$ in Sec.\ref{subsec:ormt} in the near-horizon.  
For this, we replace $r$ in those $j^{(2)}_{MN}$ with
$r_{h,4D} +\Delta r$  
($\Delta r=r-r_{h,4D}$ and $r_{h,4D}$ is given in (\ref{lhrb2}))\footnote{
We give the Jacobian and partial derivatives when we change from $r$ to $\Delta r$. 
We denote the old and new coordinates as ($r$, $\theta$) to ($\Delta r$, $\tilde{\theta}$). 
The relations between these are $\Delta r = r-r_h(\theta)$ and $\tilde{\theta}=\theta$. 
Then, 
$
drd\theta = 
\bigg|
\begin{array}{cc}
\frac{\partial r}{\partial (\Delta r)} & \frac{\partial r}{\partial \tilde{\theta}} \\
\frac{\partial \theta}{\partial (\Delta r)}  & \frac{\partial \theta}{\partial \tilde{\theta}} \\
\end{array}
\bigg|
d(\Delta r)d\tilde{\theta}=d(\Delta r)d\tilde{\theta}
$. Further, 
$
\frac{\partial}{\partial r}
=\frac{\partial (\Delta r)}{\partial r}\frac{\partial}{\partial (\Delta r)}
+\frac{\partial \tilde{\theta}}{\partial r}\frac{\partial}{\partial \tilde{\theta}}
=\frac{\partial}{\partial (\Delta r)}
$
and 
$
\frac{\partial}{\partial \theta}
=\frac{\partial (\Delta r)}{\partial \theta}\frac{\partial}{\partial (\Delta r)}
+\frac{\partial \tilde{\theta}}{\partial \theta}\frac{\partial}{\partial \tilde{\theta}}
=
-\frac{\partial r_h(\theta)}{\partial \theta}\frac{\partial}{\partial (\Delta r)}
+\frac{\partial}{\partial \tilde{\theta}}. 
$
}, 
then expand around $\Delta r=0$. 
Writing these as $t_{MN}$,  
\newpage
\begin{eqnarray}
\label{ctksttnh} 
t_{tt} 
\!\!\!\! &=& \!\!\!\!
\Big(
-\frac{\Delta r}{2 m}+\frac{\Delta r^2}{4 m^2}+O\left(\Delta r^3\right)
\Big)
+\varepsilon ^2 
\frac{15\sin ^2(2 \theta )}{4\pi}
\Big\{
\frac{3  \Delta r}{4 m }
-\frac{\sqrt{2}  \Delta r^{3/2}}{m^{3/2}  }
+\frac{3 \Delta r^2}{4 m^2  }
\nonumber\\*
&&
+\frac{  \Delta r^{5/2}}{2\sqrt{2} m^{5/2} }
+O\left(\Delta r^3\right)
\Big\}
+O(\varepsilon ^3), 
\end{eqnarray}
\begin{eqnarray}
\label{ctksrrnh} 
t_{rr} 
\!\!\!\! &=& \!\!\!\!
\Big(
\frac{2 m}{\Delta r}+1+O\left(\Delta r^3\right)
\Big)
+\varepsilon ^2 \frac{15\sin ^2(2\theta )}{\pi}
\Big\{
\frac{3 m }{4 \Delta r}
-\frac{\sqrt{2m} }{ \sqrt{\Delta r}}
+\frac{6}{4}
- \frac{3 \sqrt{\Delta r}}{2\sqrt{2m} }
+\frac{2 \Delta r}{4m}
\nonumber\\*
&&
- \frac{3 \Delta r^{3/2}}{16 \sqrt{2} m^{3/2} }
+ O\left(\Delta r^{5/2}\right)
\Big\}
+O(\varepsilon ^3),
\end{eqnarray}
\begin{eqnarray}
\label{ctksthnh}  
t_{\theta\theta} 
\!\!\!\! &=& \!\!\!\!
\Big(
4 m^2+4 \Delta r m+\Delta r^2+O\left(\Delta r^3\right)
\Big)
+\varepsilon 24 \sqrt{\frac{5}{\pi }} \cos (2 \theta )
\Big\{
m^2-m^{3/2} \sqrt{2} \sqrt{\Delta r}
\nonumber\\*
&&
+2 m  \Delta r
-\frac{5}{4} \sqrt{2m}  \Delta r^{3/2}
+\frac{5}{4} \Delta r^2
+O\left(\Delta r^{5/2}\right) 
\Big\}
+
\frac{90\varepsilon ^2}{\pi }
\Big\{
2m^2 \cos^2 (2 \theta )
\nonumber\\*
&&
-4 \sqrt{2} m^{3/2} \cos ^2(2 \theta )\sqrt{\Delta r}
+\frac{m (41 \cos (4 \theta )+39) \Delta r}{8}
-\frac{\sqrt{2m}}{3} (14 \cos (4 \theta )+13) \Delta r^{3/2}
\nonumber\\*
&&
+\frac{5 (21 \cos (4\theta )+19) \Delta r^2}{16}
+O\left(\Delta r^{5/2}\right)
\Big\}
+O\left(\varepsilon^3\right),
\end{eqnarray}
\begin{eqnarray}
\label{ctksppnh} 
t_{\phi\phi}
\!\!\!\! &=& \!\!\!\! 
\sin ^2(\theta )
\left(4 m^2 
+4 \Delta r m
+\Delta r^2 
+O\left(\Delta r^3\right)\right)
+
3 \varepsilon 
\sqrt{\frac{5}{\pi }} \sin ^2(2 \theta )  
\Big\{
2 m^2
-2 \sqrt{2} m^{3/2}  \sqrt{\Delta r}
\nonumber\\*
&&
+4 m  \Delta r 
-5 \sqrt{m} \Delta r^{3/2}
+\frac{3}{2} \Delta r^2
+O\left(\Delta r^{5/2}\right)
\Big\}
+
\varepsilon ^2
\frac{45 \sin ^2(2 \theta )}{\pi }
\Big\{
m^2 \cos (2 \theta )
\nonumber\\*
&&
-2\sqrt{2} m^{3/2} \cos (2 \theta ) \sqrt{\Delta r}
+ \frac{m}{4}  (21 \cos (2 \theta )-1) \Delta r
+\frac{5 (11 \cos (2 \theta )-1) \Delta r^2}{8} 
\nonumber\\*
&&
+\frac{675}{2} \sqrt{2m}  (2-29 \cos (2 \theta )) \Delta r^{3/2}
+O\left(\Delta r^{5/2}\right)
\Big\}
+O(\varepsilon ^3),
\end{eqnarray}
\begin{eqnarray}
\label{ctksrhnh} 
t_{r\theta} 
\!\!\!\! &=& \!\!\!\!
-3\varepsilon  
\sqrt{\frac{5}{\pi}}
\sin (2 \theta) 
\Big\{
\frac{m^{3/2} \sqrt{2}}{\sqrt{\Delta r}}
-2 m 
+\frac{7}{2}\sqrt{\frac{m}{2}} \sqrt{\Delta r}
- 6  \Delta r
+\frac{27 \Delta r^{3/2}}{16 \sqrt{2m}}
-\frac{\Delta r^2}{2m}
\nonumber\\* 
&&
+O\left(\Delta r^{5/2}\right)
\Big\}
+\varepsilon^2 
\frac{15\sin (4 \theta )}{\pi} 
\frac{m^2 }{2   \Delta r}
-\frac{3 m^{3/2}}{\sqrt{2}  \sqrt{\Delta r}}
+\frac{19 m}{4  }
-\frac{41 \sqrt{m}\sqrt{\Delta r}}{4 \sqrt{2}  }
\nonumber\\*
&&
+\frac{33 \Delta r}{4 }
-\frac{329\Delta r^{3/2}}{32\sqrt{2m} }
+\frac{5 \Delta r^2}{m }
+O\left(\Delta r^{5/2}\right)
\Big\}
+O(\varepsilon ^3). 
\end{eqnarray}
\newpage
Contravariant  metrices toward these are obtained as  
\begin{eqnarray}
\label{ctksttnu}
t^{tt} 
\!\!\!\! &=& \!\!\!\!
\Big(
-\frac{2 m}{\Delta r}-1-\frac{\Delta r}{2m}+O\left(\Delta r^2\right)
\Big)
-
\varepsilon ^2 
 \frac{45\sin ^2(2 \theta )}{2\pi}
\Big(
\frac{m}{2 \Delta r}
-\frac{ 2\sqrt{2m} }{ 3\sqrt{\Delta r}}
+1
-\frac{\sqrt{\Delta r}}{\sqrt{2m}   }
\nonumber\\*
&&
+\frac{7 \Delta r}{8m}
+O\left(\Delta r^{3/2}\right)
\Big)
+O(\varepsilon ^3),
\end{eqnarray}
\begin{eqnarray}
\label{ctksrrnu}
t^{rr} 
\!\!\!\! &=& \!\!\!\!
\Big(
\frac{\Delta r}{2 m}+O\left(\Delta r^2\right)
\Big)
+\varepsilon ^2 \frac{15\sin ^2(2\theta )}{\pi}
\Big(
\frac{3  \Delta r}{16 m }
-\frac{  \Delta r^{3/2}}{\sqrt{2} m^{3/2} }
+O\left(\Delta r^2\right)
\Big) 
+O\left(\varepsilon^3\right),
\\  
\label{ctksthnu} 
t^{\theta\theta} 
\!\!\!\! &=& \!\!\!\!
\Big(
\frac{1}{4 m^2}
-\frac{\Delta r}{4m^3}
+O\left(\Delta r^2\right)
\Big)
+ 3 \varepsilon \sqrt{\frac{5}{\pi }} \cos (2 \theta)
\Big(
-\frac{1}{2 m^2}
+\frac{ \sqrt{\Delta r}}{\sqrt{2}m^{5/2}}
+O\left(\Delta r^{3/2}\right)
\Big)
\nonumber\\*
&&
+\varepsilon ^2 \frac{45}{8\pi}
\Big(
\frac{11 \cos(4 \theta )+13}{4m^2}
-\frac{(11 \cos (4 \theta )+13)\sqrt{\Delta r}}{ \sqrt{2} m^{5/2}}
+\frac{(65 \cos (4 \theta )+79) \Delta r}{8m^3 }
\nonumber\\*
&&
+O\left(\Delta r^{3/2}\right)
\Big)
+O(\varepsilon ^3),
\\  
\label{ctksrhnu}
t^{r\theta} 
\!\!\!\! &=& \!\!\!\!
\varepsilon  
\frac{3}{4}
\sqrt{\frac{5}{\pi}} \sin (2 \theta)
\Big(
\frac{   \sqrt{\Delta r}}{\sqrt{2}m^{3/2}}
-\frac{\Delta r}{ m^2}
+O\left(\Delta r^{3/2}\right)
\Big)
\nonumber\\*
&&
-\varepsilon ^2 \frac{15\sin (4\theta )}{16 \pi} 
\Big(
\frac{1}{m}
+\frac{6 \sqrt{\Delta r}}{ \sqrt{2} m^{3/2}}
-\frac{16 \Delta r}{m^2}
+O\left(\Delta r^{3/2}\right)
\Big)
+O(\varepsilon ^3), 
\\   
\label{ctksppnu}
t^{\phi\phi} 
\!\!\!\! &=& \!\!\!\!
\frac{\csc ^2\theta }{4}
\Big(
\frac{1}{ m^2}
-\frac{\Delta r}{m^3}
+O\left(\Delta r^2\right)
\Big)
+3 \varepsilon  \sqrt{\frac{5}{2\pi }} \cot ^2\theta 
\Big(
-\frac{1}{\sqrt{2}m^2}
+\frac{\sqrt{\Delta r}}{  m^{5/2}}
+O\left(\Delta r^{3/2}\right)
\Big)
\nonumber\\*
&&+\varepsilon ^2  \frac{45\cot ^2(\theta )}{\pi} 
\Big(
\frac{ \cos (2\theta )+2}{4 m^2}
-\frac{(\cos(2 \theta )+2) \sqrt{\Delta r}}{\sqrt{2} m^{5/2}}
+\frac{(11 \cos (2 \theta )+25) \Delta r}{16 m^3}
\nonumber\\*
&& 
+O\left(\Delta r^{3/2}\right)
\Big)
+O(\varepsilon ^3).
\end{eqnarray}
We can check these  are the inverse  each other in the range of $\varepsilon^2$. 
(Leading of these are the same with just a Schwarzschild, 
which is the technical reason for our result, no change.)  

\subsection{Hawking temperature in the original 4D}
\label{subsec:hwkkdk} 

We have given the Killing vector in (\ref{subsec:ormt}), 
and obtained the position of the horizon in (\ref{lhrb2}) and the near-horizon metrices.  
With these and the formula: 
$\kappa^2=-\frac{1}{2}D^M \xi^N D_M \xi_N$,
the Hawking temperature in the original 4D spacetime can be obtained as 
\begin{eqnarray}
\label{hkod4d}
T_H ={1}/{8\pi m}+O(\varepsilon ^3). 
\end{eqnarray}
This is the same with the one in just the Schwarzschild. 
We can understand this as follows.
 
Generally, $T_H = \frac{1}{4\pi}\big|\partial_r f(r) \big|_{r=r_h}\big|$ 
for $ds^2=-f(r)dt^2+f^{-1}(r)dr^2+\cdots$, 
where $f(r)|_{r=r_h}=0$ (these $f(r)$ and $r_h$ are irrelevant with this study). 
However, our $f(r)$ behaves same with just the Schwarzschild at $\Delta r=0$ ($r=r_{h,4D}$) as in (\ref{ctksttnh}). 

Our Hawking temperature might have been expected to depend on the angular directions, 
which breaks the zeroth law of black hole thermodynamics. 
However, we could expect from the result above that 
it would be always out of the analysis's order. 
 
\subsection{Near-horizon action}
\label{subsec:nha}

Having obtained 4D near-horizon metrices, 
let us obtain near-horizon action. 
For this, we write down (\ref{sacf}) term by term, 
then express each line order by order as to $\varepsilon ^2$-order as
\begin{subequations}\label{acecl} 
\begin{align}
\label{sfaa1} 
\textrm{$\cal L$ of (\ref{sacf})}=
-& 
\phi^* \partial_t 
\big\{
\big( \sqrt{-t}^{(0)}+\sqrt{-t}^{(1)}+\sqrt{-t}^{(2)} \big) \big(t^{tt}{}^{(0)}+t^{tt}{}^{(2)}\big) \partial_t 
\big\} \phi\\*
\label{sfaa2}
-&
\phi^* \partial_r 
\big\{
\big(\sqrt{-t}^{(0)}+\sqrt{-t}^{(1)}+\sqrt{-t}^{(2)}\big) \big(t^{rr}{}^{(0)}+t^{rr}{}^{(2)}\big) \partial_r 
\big\} \phi\\*
\label{sfaa3}
-&
\phi^* \partial_r
\big\{ 
\big(\sqrt{-t}^{(0)}+\sqrt{-t}^{(1)}+\sqrt{-t}^{(2)}\big) \big(t^{r\theta}{}^{(1)}+t^{r\theta}{}^{(2)} \big) \partial_\theta
\big\}\phi\\*
\label{sfaa4}
-&
\phi^* \partial_\theta 
\big\{ \big(\sqrt{-t}^{(0)}+\sqrt{-t}^{(1)}+\sqrt{-t}^{(2)}\big) 
\big(t^{\theta\theta}{}^{(0)}+t^{\theta\theta}{}^{(1)}+t^{\theta\theta}{}^{(2)}\big)  \partial_\theta
\big\}\phi\\*
\label{sfaa5}
-&
\phi^* \partial_\theta
\big\{ 
\big(\sqrt{-t}^{(0)}+\sqrt{-t}^{(1)}+\sqrt{-t}^{(2)}\big) \big(t^{r\theta}{}^{(1)}+t^{r\theta}{}^{(2)}\big)  \partial_r
 \big\}\phi\\*
\label{sfaa6}
-&
\phi^* \partial_\phi \big\{
\big(\sqrt{-t}^{(0)}+\sqrt{-t}^{(1)}+\sqrt{-t}^{(2)}\big) \big(t^{\phi\phi}{}^{(0)}+t^{\phi\phi}{}^{(1)}+t^{\phi\phi}{}^{(2)}\big) \partial_\phi
\big\}\phi,
\end{align}
\end{subequations}
where the numbers in the superscripts mean the part of that quantity at that order when that quantity is 
expanded with regard to $\varepsilon$~\footnote{
E.g., $\sqrt{-t}^{(0)}+\sqrt{-t}^{(1)}+\sqrt{-t}^{(2)}$ 
means the first three terms writing as $\sqrt{-t}=(\cdots)+(\cdots)\varepsilon+(\cdots)\varepsilon^2+\cdots$.
}.

We write the order behavior of the ingredients in (\ref{sfaa1})-(\ref{sfaa6}) based on (\ref{ctksttnu})-(\ref{ctksppnu}) as
\begin{itemize}
\item $\displaystyle \sqrt{-t}^{(0)}+\sqrt{-t}^{(1)}+\sqrt{-t}^{(2)} \sim (1+\Delta r)+(1+\sqrt{\Delta r})\varepsilon+(1+\sqrt{\Delta r})\varepsilon^2$,
\item $\displaystyle t^{tt}{}^{(0)}+t^{tt}{}^{(2)} \sim (\Delta r)^{-1}+\varepsilon^2 (\Delta r)^{-1}$,
\item $\displaystyle t^{rr}{}^{(0)}+t^{rr}{}^{(2)} \sim \Delta r+\Delta r \varepsilon^2$,
\item $\displaystyle t^{r\theta}{}^{(1)}+t^{r\theta}{}^{(2)} \sim \sqrt{\Delta r}\varepsilon+(1+\sqrt{\Delta r})\varepsilon^2$,
\item $\displaystyle t^{\theta\theta}{}^{(0)}+t^{\theta\theta}{}^{(1)}+t^{\theta\theta}{}^{(2)} \sim (1+\Delta r)+(1+\sqrt{\Delta r})\varepsilon+(1+\sqrt{\Delta r})\varepsilon^2$,
\item $\displaystyle t^{\phi\phi}{}^{(0)}+t^{\phi\phi}{}^{(1)}+t^{\phi\phi}{}^{(2)} \sim (1+\Delta r)+(1+\sqrt{\Delta r})\varepsilon+(1+\sqrt{\Delta r})\varepsilon^2$.
\end{itemize}
With these, we can get the order behavior of the each line (\ref{sfaa1})-(\ref{sfaa6}) as
\begin{eqnarray}
\label{acnh1}
\textrm{(\ref{sfaa1})} 
\!\!\!\! &\sim& \!\!\!\!
\Big(\frac{1}{\Delta r}+1+O\left(\Delta r^2\right)\Big)
+\varepsilon  
\Big(\frac{1}{\Delta r}+\frac{1}{\sqrt{\Delta r}}+O\left(\Delta r^{3/2}\right)\Big)
\nonumber\\*
&& 
+\varepsilon ^2
\Big(\frac{2}{\Delta r}+\frac{1}{\sqrt{\Delta r}}+1+O\left(\Delta r^{3/2}\right)\Big)
+O(\varepsilon ^3),
\nonumber\\ 
\label{acnh2}
\textrm{(\ref{sfaa2})}
\!\!\!\! &\sim& \!\!\!\!
\Big(1+2 \Delta r+O\left(\Delta r^2\right)\Big)
+\varepsilon  
\Big(
1+\frac{3\sqrt{\Delta r}}{2}+O\left(\Delta r^{3/2}\right)
\Big)
\nonumber\\*
&&
+\varepsilon ^2 
\Big(
2
+\frac{3\sqrt{\Delta r}}{2}
+2 \Delta r
+O\left(\Delta r^{3/2}\right)
\Big)
+O(\varepsilon ^3),
\nonumber\\ 
\label{acnh3}
\textrm{(\ref{sfaa3})}
\!\!\!\! &\sim& \!\!\!\!
\varepsilon 
\Big(
\frac{1}{2 \sqrt{\Delta r}}+\frac{3\sqrt{\Delta r}}{2}+O\left(\Delta r^{3/2}\right)
\Big)
+
\varepsilon ^2
\Big(
\frac{1}{\sqrt{\Delta r}}
+2
+\frac{3\sqrt{\Delta r}}{2}
+O\left(\Delta r^{3/2}\right)
\Big)
+O(\varepsilon ^3),
\nonumber
\\ 
\label{acnh4}
\textrm{(\ref{sfaa4})}
\!\!\!\! &\sim& \!\!\!\!
\big(
1+2 \Delta r
+O\big(\Delta r^2\big)
\big)
+\varepsilon  
\big(
2
+2 \sqrt{\Delta r}
+2 \Delta r
+O\big(\Delta r^{3/2}\big)
\big)
\nonumber\\*
&&
+\varepsilon ^2 
\big(
3
+4 \sqrt{\Delta r}
+3 \Delta r
+O\big(\Delta r^{3/2}\big)
\big)
+O\big(\varepsilon ^3\big),
\nonumber
\\ 
\label{acnh5}
\textrm{(\ref{sfaa5})}
\!\!\!\! &\sim& \!\!\!\!
\varepsilon  
\big(
\sqrt{\Delta r}
+O\big(\Delta r^{3/2}\big)
\big)
+\varepsilon ^2 
\big(
1
+2\sqrt{\Delta r}
+2 \Delta r
+O\big(\Delta r^{3/2}\big)
\big)
+O\big(\varepsilon ^3\big),
\nonumber
\\ 
\label{acnh6} 
\textrm{(\ref{sfaa6})}
\!\!\!\! &\sim& \!\!\!\!
\big(
1
+2 \Delta r
+O\big(\Delta r^2\big)
\big)
+\varepsilon  
\big(
2
+2 \sqrt{\Delta r}
+2 \Delta r
+O\big(\Delta r^{3/2}\big)
\big)
\nonumber\\*
&&
+\varepsilon ^2 
\big(
3
+4\sqrt{\Delta r}
+3 \Delta r
+O\big(\Delta r^{3/2}\big)
\big)
+O\big(\varepsilon ^3\big).
\nonumber
\end{eqnarray}

We find (\ref{sfaa1}) is dominant and others are vanish or ignorable compared with (\ref{sfaa1}) at $\Delta r \to 0$. 
Therefore, from the viewpoint of which parts remain at $\Delta r \to 0$, we may remain only (\ref{sfaa1}). 
However, since the ($t$, $r$)-space is crucial in the analysis of Hawking temperature and flux,  
the parts on and tangling with the ($t$, $r$)-space are indispensable in our analysis.   
(Also note that (\ref{sfaa2}) makes the action at $\Delta r \to 0$ finite.) 
Therefore, remaining (\ref{sfaa2}), (\ref{sfaa3}) and (\ref{sfaa5}) in addition to (\ref{sfaa1}), we consider the following action: 
\begin{eqnarray}\label{acnkt}
\textrm{(\ref{acecl})}
\!\!\!\! &=& \!\!\!\!
\sqrt{-t}
(
t^{tt}\partial_t \phi^* \partial_t \phi 
- t^{rr}\partial_r \phi^* \partial_r \phi
- t^{r \theta}\partial_r \phi^* \partial_\theta \phi
- t^{\theta r}\partial_\theta \phi^* \partial_r \phi
) + \cdots \nonumber\\*
\!\!\!\! &=& \!\!\!\!
\sqrt{-t}
\big(
t^{tt}\partial_t \phi^* \partial_t \phi
+t^{rr}
\big(\partial_r  \phi^* + \frac{t^{\theta r}}{t^{rr}} \partial_\theta\phi^* \big)
\big(\partial_r  \phi + \frac{t^{\theta r}}{t^{rr}}\partial_\theta\phi \big)  
-\frac{(t^{\theta r})^2}{t^{rr}} \partial_\theta \phi^* \partial_\theta \phi
\big), 
\end{eqnarray} 
as the near-horizon action, where ``$\cdots$'' is (\ref{sfaa4}), (\ref{sfaa6}) 
(the terms vanishing at $\Delta r \to 0$ and irrelevant for the ($t$, $r$)-space) 
and terms under $\varepsilon^3$-order.

Let us look ${(t^{\theta r})^2}/{t^{rr}}$ and ${t^{\theta r}}/{t^{rr}}$ in (\ref{acnkt}). 
Using (\ref{ctksttnu})-(\ref{ctksppnu}), we can write these as 
\begin{subequations}
\begin{align}
\label{kkrd}
\bullet 
\quad&
\hspace{5.5mm}
\frac{t^{\theta r}}{t^{rr}} 
=  
\varepsilon  
\frac{3}{2}\sqrt{\frac{5}{\pi }}\sin (2 \theta )
\Big(
 \frac{1}{\sqrt{2m} \sqrt{\Delta r}}
-\frac{1}{m}
+O\left(\sqrt{\Delta r}\right)
\Big)
\nonumber
\\*
&
\hspace{15.0mm}
-\varepsilon ^2 \frac{15}{8\pi} \sin (4 \theta )
\Big(
\frac{1}{\Delta r}
+\frac{6}{\sqrt{2m} \sqrt{\Delta r}}
+O\left(\Delta r^0\right)
\Big)
+O(\varepsilon ^3),
\\*
\label{kkew}
\bullet 
\quad&
\frac{(t^{\theta r})^2}{t^{rr}} = 
\varepsilon ^2 
\frac{45 \sin ^2(2 \theta )}{16\pi}
\Big(
\frac{1}{m^2}
-\frac{4\sqrt{\Delta r}}{\sqrt{2} m^{5/2} }+O\left(\Delta r^1\right)
\Big)
+O(\varepsilon ^3).   
\end{align}
\end{subequations}
Thus, ${(t^{\theta r})^2}/{t^{rr}}$ is ignorable in the limit $\Delta r \to 0$, but 
${t^{\theta r}}/{t^{rr}}$ is not. 
However ${t^{\theta r}}/{t^{rr}}$ is ignorable finally 
in the analysis of Hawking temperature and flux for the following facts: 
\begin{itemize}
\item     
We can regard $\frac{t^{\theta r}}{t^{rr}}\partial_\theta\phi$ 
as the $r$-component of U(1) gauge field 
in the sense that 
we can evaluate the anomalies and currents associated with it 
using the way to evaluate those for U(1) gauge field. 
The point here is that it is composed of the $t$-independent $r$-component only, 
therefore  we can see by looking at (4) in \cite{Iso:2006wa} 
the gauge anomalies do not arise 
from $\frac{t^{\theta r}}{t^{rr}}\partial_\theta\phi$. 
Hence, $\frac{t^{\theta r}}{t^{rr}}\partial_\theta\phi$ is irrelevant of this study. 
\item 
Next, as for the gravitational anomalies, 
since $J^\mu$ is zero according to (4) in \cite{Iso:2006wa}, 
the second term in r.h.s. in (16) in \cite{Iso:2006wa} is some constants. 
The first term in the (16) will be also zero, 
since our gauge field is composed of only $t$-independent  $r$-component. 
\end{itemize}
Hence, since ${t^{\theta r}}/{t^{rr}}$ has nothing to do with gauge and gravitational anomalies, 
we can ignore it in our analysis 
and are allowed to write the near-horizon action we consider as
\begin{eqnarray}\label{cncwra}
S_{\rm nh} = \frac{1}{2}\int dx^4 \sqrt{-t} (t^{tt} \partial_t  \phi^* \partial_t \phi +t^{rr} \partial_r \phi^* \partial_r \phi ). 
\end{eqnarray}

Let us obtain the concrete expression for (\ref{cncwra}). For this we write as
\begin{subequations}
\label{acaftl}
\begin{align}
\label{acaftl1}
\textrm{$\cal L$ of (\ref{cncwra})}=
&
-
\phi^*
\{ 
(\sqrt{-t})^{(0)} t^{tt}{}^{(0)}
\}
\partial_t \partial_t \phi 
-
\phi^*\partial_r 
\{ 
(\sqrt{-t})^{(0)} t^{rr}{}^{(0)}\partial_r 
\phi
\}
\\*
\label{acaftl2}
&
-\phi^*
\{
(\sqrt{-t})^{(0)} t^{tt}{}^{(2)}
+
(( \sqrt{-t} )^{(1)}+(\sqrt{-t})^{(2)})
g^{tt}{}^{(0)}
\}
\partial_t \partial_t \phi 
\\*
\label{acaftl3}
&
-\phi^*
\partial_r
\{
((\sqrt{-t})^{(0)} t^{tt}{}^{(2)}+((\sqrt{-t})^{(1)}+(\sqrt{-t})^{(2)})
t^{tt}{}^{(0)}
)
\partial_r \phi
\},  
\end{align}
\end{subequations}
where the meaning of the numbers in the superscripts are the same with (\ref{acecl}).

We list the ingredients needed to calculate (\ref{acaftl}) as
\begin{subequations}
\label{egacal}
\begin{align}
\label{egacal1}
&
\bullet \quad (\sqrt{-t})^{(0)} 
=4 m \sin (\theta ) (\Delta r+m),\\*
&
\bullet \quad  \sum_{i=1,2}(\sqrt{-t})^{(i)} 
= 
\frac
{90 \varepsilon ^2 m^2 \sin (\theta ) \cos ^2(\theta ) (3 \cos (2\theta )-1)}{\pi }
+6 \sqrt{\frac{5}{\pi }} \varepsilon  m^2 \sin (\theta) (3 \cos (2 \theta )+1),\\*
\label{egacal2}
&
\bullet \quad 
t^{tt}{}^{(0)} = -{2 m}/{\Delta r},\quad 
t^{rr}{}^{(0)} = {\Delta r}/{2 m},
\\*
\label{egacal3}
&
\bullet \quad t^{tt}{}^{(2)}  = 
-{45 \varepsilon ^2 m \sin ^2(2 \theta )}/{4 \pi  \Delta r},\quad
t^{rr}{}^{(2)}  =  
{45 \Delta r \varepsilon ^2 \sin ^2(2\theta )}/{16 \pi  m}.
\end{align}
\end{subequations}
Using these, we can obtain the concrete expression of  (\ref{cncwra}) to $\varepsilon^2$-order as
\begin{eqnarray}\label{acale2}
\textrm{(\ref{cncwra})} 
\!\!\! &=& \!\!\!
- \int d^4x\,  (2m)^2 \sin (\theta ) 
\Big\{
1
+\frac{3}{2} \sqrt{\frac{5}{\pi }} \left(1 +3 \cos (2 \theta ) \right) \varepsilon
+\frac{45}{2\pi} 
\Big(
\frac{ \sin (2 \theta)}{4} 
-\cos ^2(\theta )
\nonumber \\* 
&&
+3  \cos ^2(\theta ) \cos (2 \theta )
\Big) \varepsilon ^2
\Big\}
\phi^* (t^{tt}\partial_t\partial_t+\partial_r(t^{rr}\partial_r))\phi
\\
\label{2defmt}
(t^{tt},t^{rr})
\!\!\! &=& \!\!\!
\Big(-\frac{2m}{\Delta r},\,\frac{\Delta r}{2m}\Big)
\nonumber \\*
\!\!\! &=& \!\!\!
\Big(
-\frac{2 m}{r-2 m}
-\frac{15 \varepsilon ^2 m^2 \sin ^2(2 \theta )}{4 \pi  (r-2 m)^2},\,
\frac{r-2 m}{2 m}-\frac{15 \varepsilon ^2 \sin ^2(2 \theta )}{16 \pi }
\Big).
\end{eqnarray} 
Note that the mass parts 
in the denominator of $-\frac{2m}{\Delta r}$ and numerator of $\frac{\Delta r}{2m}$
do not agree each other. 
We fix  this point in (\ref{defgtt}) by defining the effective mass (\ref{mdfeft}).

\subsection{Integrate out of ($\theta$, $\phi$)}
\label{subsec:nhar}

We will obtain the 2D effective near-horizon action 
by integrating out  ($\theta$, $\phi$) of (\ref{acale2}).
For this, we first expand $\phi$ by the spherical harmonics as
\begin{eqnarray}\label{pexsf}
\phi(t,r,\theta,\phi)=\sum_{l,m}\phi_{lm}(t,r)Y_l^m(\theta,\phi).
\end{eqnarray}
Then defining the following $\Lambda_{lm,\,kn}$ ($d\Omega=d\theta d\phi \sin \theta$), we can write (\ref{acale2}) as
\begin{subequations}
\begin{align}
\label{nvowr}
\Lambda_{lm,\,kn} 
=& 
\int d\Omega
\Big\{
1+\frac{3}{2} \sqrt{\frac{5}{\pi }} \left(1 +3 \cos (2 \theta ) \right) \varepsilon
\nonumber\\* 
&
+\frac{45}{2\pi}  
\Big(
\frac{ \sin (2 \theta)}{4} -\cos ^2(\theta )+3  \cos ^2(\theta ) \cos (2 \theta )
\Big) \varepsilon ^2
\Big\} 
\, (Y_l^m)^*Y_k^n,
\\
\label{ef2dacp}
{\rm (\ref{acale2})}
=& 
-\sum_{l, m} \sum_{k, n} \int dt dr (2m)^2  
\Big\{
\nonumber\\*
&
\hspace{5.0mm}
\phi_{lm}^* \! 
\Big( 
-\frac{2m}{r-2m}\Lambda_{lm,\,kn} 
-\frac{15m^2\varepsilon^2}{4\pi(r-2m)^2}
\! \int \!  
d\Omega \sin^2 (2\theta)
(Y_l^m)^* Y_k^n
\Big) 
 \partial_t \partial_t \phi_{kn}
\nonumber\\*
&
+\phi_{lm}^* \partial_r \!
\Big(
\frac{r-2m}{2m}\Lambda_{lm,\,kn} 
-\frac{15m^2\varepsilon^2}{16\pi}
\! \int \! d\Omega\sin^2 (2\theta)
(Y_l^m)^* Y_k^n
\Big)  
 \partial_r\phi_{kn}
\Big\}.
\end{align}
\end{subequations}

We can evaluate all kinds of the ($\theta$, $\phi$)-integrals in (\ref{ef2dacp}) (totally four) using (\ref{fmry3}) as\footnote{
Necessary formulas for the calculations  (\ref{intp1})-(\ref{intp4}):
\begin{equation} 
\label{fmry3}
\begin{split}
&(Y_l^m)^*=(-)^m Y_l^{-m} , \quad 
\int d\Omega~  (Y_{l_1}^{m_1})^* Y_{l_2}^{m_2}=\delta_{l_1 l_2} \delta_{m_1 m_2}, \quad
\int d\Omega~ \sin 2\theta (Y_{l_1}^{m_1})^* Y_{l_2}^{m_2}=0 \quad {\rm and}
\\
&\int d\Omega~ (Y_{l_1}^{m_1})^* (Y_{l_2}^{m_2})^* Y_L^M=\sqrt{\frac{(2l_1+1)(2l_2+1)}{4\pi (2L+1)}}
\langle l_1 \, 0 \, l_2 \, 0| L \, 0 \rangle \langle l_1 \, m_1~l_2 \, m_2| L \, M \rangle,
\end{split}
\end{equation}
where $\langle l_1 m_1 l_2 m_2| LM \rangle$ mean Clebsch-Gordan coefficients \cite{wikiy3}. 
We can obtain (\ref{intp1})-(\ref{intp4}) using (\ref{fmry3}) by rewriting these integrands into the form of the 3 products 
of spherical harmonics. To be concrete, express 
$\cos (2 \theta)$, 
$\cos^2  \theta$,
$\sin^2 (2 \theta)$ and 
$\cos^2  \theta \cos (2 \theta)$ 
by $Y_0^0$, $Y_2^0$ and $Y_4^0$ (e.g. $\cos (2 \theta)=\frac{8}{3}\sqrt{\frac{\pi}{5}}Y_2^0-\frac{2\sqrt{\pi}}{3}Y_0^0$).
}   
\begin{subequations}
\begin{align}
\label{intp1}
\bullet  \quad & 
\int d\Omega \cos 2\theta \,(Y_k^n)^* Y_l^m 
\nonumber\\*
=& \,\,
-\frac{4 m^2-1} {4 l^2+4 l-3}\delta _{kl}\delta _{nm}
+\frac{2 (-1)^{2 m} 
\sqrt{\frac{((l+1)^2-m^2) ((l+2)^2-m^2)}{(2 l+1) (2 l+5)}} }{2 l+3}\delta _{k-2,l}\delta _{nm} 
\nonumber\\*  
\equiv& \,\, 
{\cal I}^{A_0}_{lm}\,\delta _{kl} \delta _{nm}+{\cal I}^{A_2}_{lm}\,\delta _{k-2,l} \delta _{nm}
\equiv{\cal I}^{A}_{lm,\,kn}, 
\\[1.0mm]
\label{intp2}
\bullet  \quad &
\int d\Omega \cos^2 \theta \,(Y_k^n)^* Y_l^m 
\nonumber\\* 
=& \,\,
\frac{2 l^2+2 l-2 m^2-1}{4 (l+1)^2-4 (l+1)-3}\delta _{kl}\delta _{nm}
+\frac{(-1)^{2 m} \sqrt{\frac{((l+1)^2-m^2) ((l+2)^2-m^2)}{(2 l+1) (2 l+5)}} }{2 l+3}\delta _{k-2,l}\delta _{nm} 
\nonumber\\*  
\equiv& \,\, 
{\cal I}^{B_0}_{lm}\,\delta _{kl} \delta _{nm}+{\cal I}^{B_2}_{lm} \,\delta _{k-2,l}\delta _{nm}
\equiv{\cal I}^{B}_{lm,\,kn}, 
\\[1.0mm]
\label{intp3}
\bullet  \quad &
\int d\Omega \sin^2 (2\theta) \, (Y_k^n)^* Y_l^m 
& \nonumber\\*
=& \,\, \frac{8 (-1)^{2 m}  (l (l+1)(l^2+l-5)+2 l (l+1) m^2-3 m^4+3)}{(2 l-3) (2 l-1) (2 l+3) (2 l+5)} \delta _{kl} \delta _{nm}
\nonumber\\*
&+4 (-1)^{2 m-2 l} \sqrt{\frac{((l+1)^2-m^2) ((l+2)^2-m^2)}{4 l^2+12 l+5}} 
\nonumber\\*
& \times \frac{(-1)^{2 l} (2 l-1) (2 l+7)-4 (-1)^{2 m} (l (l+3)-7 m^2) }{7 (2 l-1) (2 l+3) (2 l+7)} \delta _{k-2,l}\delta _{nm}
\nonumber\\*
&-4 (-1)^{2m} 
\sqrt{\frac{((l+1)^2-m^2) ((l+2)^2-m^2) ((l+3)^2-m^2) ((l+4)^2-m^2) }{(2 l+1) (2 l+3)^2 (2 l+5)^2 (2 l+7)^2 (2 l+9)}} \delta _{k-4,l}\delta _{nm} 
\nonumber\\*  
\equiv& \,\,  
{\cal I}^{C_0}_{lm}\,\delta _{kl} \delta _{nm}+{\cal I}^{C_2}_{lm} \,\delta _{nm} \delta _{k-2,l}+{\cal I}^{C_4}_{lm} \,\delta _{nm} \delta _{k-4,l}
\equiv{\cal I}^{C}_{lm,\,kn}, 
\\[1.0mm]
\label{intp4}
\bullet  \quad &
\int d\Omega \cos 2\theta \cos^2 \theta\, (Y_k^n)^* Y_l^m 
& \nonumber\\*
=& \,\, \frac{(-1)^{2 m} (2 (-8 l (l+1) m^2+l(l+1) (2 l (l+1)-7)+6 m^4)+30 m^2+3) }{(2 l-3) (2 l-1) (2 l+3) (2 l+5)} \delta _{kl} \delta _{nm}
\nonumber\\*
&+(-1)^{2 m-2 l} \sqrt{\frac{((l+1)^2-m^2) ((l+2)^2-m^2)}{4 l^2+12 l+5}} 
\nonumber\\*
&\times\frac{ 8 (-1)^{2 m} (l (l+3)-7 m^2)+5 (-1)^{2 l} (2 l-1) (2 l+7)}{7 (2 l-1) (2 l+3) (2 l+7)} \delta _{nm} \delta _{k-2,l} 
\nonumber\\*
&+2 (-1)^{2m} 
\sqrt{\frac{((l+1)^2-m^2) ((l+2)^2-m^2) ((l+3)^2-m^2) ((l+4)^2-m^2) }{(2 l+1) (2 l+3)^2 (2 l+5)^2 (2 l+7)^2 (2 l+9)}} 
\delta _{nm} \delta _{k-4,l} 
\nonumber\\* 
\equiv& \, \, 
{\cal I}^{D_0}_{lm} \, \delta _{kl} \delta _{nm}+{\cal I}^{D_2}_{lm}\, \delta _{nm} \delta _{k-2,l}+{\cal I}^{D_4}_{lm} \, \delta _{nm} \delta _{k-4,l}
\equiv{\cal I}^{D}_{lm,\,kn}.
\end{align}
\end{subequations}

Next  problem is it is not diagonalized with regard to $k$ and $l$. 
This reflects the shape of the horizon of our 4D black hole is not a sphere. 
Actually, it depends on ($\theta$, $\phi$) as in (\ref{lhrb2}) 
(for zeroth law of black hole thermodynamics, see Sec.\ref{subsec:hwkkdk}). 
In the next subsection, we diagonalize these  by redefining fields, 
which corresponds to rearrange appropriate bases.

\subsection{2D effective near-horizon metrices}
\label{subsec:emnh}

Using (\ref{intp1})-(\ref{intp4}), we can write (\ref{ef2dacp}) as 
\begin{eqnarray}
\!\!\! && \!\!\!
\sum_{kn} \sum_{lm} \int dtdr\, (2m)^2
\Big\{
\hspace{8.0mm}
\phi_{kn}^* 
\Big( 
-\frac{2m}{r-2m} \Lambda_{lm,\,kn} -\frac{15m^2\varepsilon^2}{4\pi(r-2m)^2} {\cal I}^{C}_{kn,\,lm}
\Big) 
\partial_t\partial_t\phi_{lm}
\nonumber\\*
\!\!\! && \!\!\!
\hspace{38.0mm}
+\,
\phi_{kn}^* \partial_r
\Big(
\frac{r-2m}{2m} \Lambda_{lm,\,kn} -\frac{15m^2\varepsilon^2}{16\pi} {\cal I}^{C}_{kn,\,lm}
\Big)  
\partial_r\phi_{lm}
\Big\}
\nonumber\\*
\label{cagttpteg}
\!\!\! &=& \!\!\!
\sum_{kn} \sum_{lm}
\int dtdr\, (2m)^2 \Lambda_{kn,\,lm} 
((g_{\rm eff})^{tt}_{kn,\,lm}  \partial_t \phi^*_{kn} \partial_t \phi_{lm}+(g_{\rm eff})^{rr}_{kn,\,lm}  \partial_r \phi^*_{kn} \partial_r \phi_{lm}),
\end{eqnarray} 
where $(g_{\rm eff})^{tt}_{kn,\,lm} $ and $(g_{\rm eff})^{rr}_{kn,\,lm}$ 
are the 2D  effective near-horizon metrices given as
\begin{subequations}
\label{defgmtpa} 
\begin{align}
\Lambda_{kn,\,lm} 
=& \,\, 
\Big( 
1+\frac{3}{2}\sqrt{\frac{5}{\pi}}\left(1+3{\cal I}^{A_0}_{lm}\right)\varepsilon
+\frac{45}{2\pi}\left(-{\cal I}^{B_0}_{lm}+3{\cal I}^{D_0}_{lm}\right)\varepsilon^2 
\Big)
\delta_{kl}\delta_{mn}
\nonumber \\*  
&
+\frac{9}{2} 
\Big( 
\sqrt{\frac{5}{\pi}} {\cal I}^{A_2}_{lm} \varepsilon
+\frac{5}{\pi}\left( -{\cal I}^{B_2}_{lm}+3{\cal I}^{D_2}_{lm} \right)\varepsilon^2 
\Big)
\delta_{k-2,l}\delta_{mn}
+\frac{135}{2\pi} {\cal I}^{D_4}_{lm} \varepsilon^2 
\delta_{k-4,l}\delta_{mn}
\nonumber\\* 
\label{ldcep} 
\equiv& \, 
\Lambda_{lm}^{(0)} \delta_{kl}\delta_{mn} + \Lambda_{lm}^{(2)}  \delta_{k-2,l}\delta_{mn} + \Lambda_{lm}^{(4)}  \delta_{k-4,l}\delta_{mn},
\\*  
\label{defgtt} 
(g_{\rm eff})^{tt}_{kn,lm} 
\equiv& \, 
-\frac{2m}{r-2 m-\frac{{\cal I}^{C}_{kn,\,lm}}{\Lambda_{kn,\,lm}}\frac{15 m \varepsilon ^2 }{8 \pi r}}
= -\frac{2(m_{\rm eff})_{kn,\,lm}}{r-2(m_{\rm eff})_{kn,\,lm}}
+O(\varepsilon ^3), 
\\* 
\label{defgrr}  
(g_{\rm eff})^{rr}_{kn,lm}  
\equiv& \, 
-((g_{\rm eff})^{tt}_{kn,lm} ) ^{-1},
\\*  
\label{mdfeft} 
(m_{\rm eff})_{kn,\,lm}
\equiv& \, 
m+\frac{15 m}{8 \pi r}{\cal I}^{C}_{kn,\,lm}\varepsilon ^2+O(\varepsilon ^3)
\equiv m + \frac{\bar{\Delta}_{kn,\,lm} }{r}\varepsilon ^2,
\end{align}
\end{subequations} 
where 
{\bf 1)} $\frac{\varepsilon ^2}{\Lambda_{kn,\,lm}} = 1 +O(\varepsilon ^3)$ in (\ref{mdfeft}), 
{\bf 2)} we defined $\Lambda_{lm}^{(0,2,4)}$, $\bar{\Delta}_{kn,\,lm}$ and $(m_{\rm eff})_{kn,\,lm}$, and  
{\bf 3)} $(m_{\rm eff})_{kn,\,lm}$ get depended on $r$, which may be concerned. 
However metrices before the near-horizon limit in Sec\ref{subsec:ormt} satisfy Einstein equation, and  
\begin{equation}\label{sdghc}
(g_{\rm eff})^{tt}_{kn,lm} =
-1+\frac{r}{2m}
-\frac{15 \, {\cal I}^{C}_{kn,\,lm}{}}{16\,m \pi} \varepsilon^2
+\frac{225\, ({\cal I}^{C}_{kn,\,lm}{})^2}{128\,m\pi^2 r} \varepsilon^4 +O\left(\varepsilon ^5\right).   
\end{equation} 
Therefore, $r$-dependence is out of the analysis's order.   
(Hawking temperature and flux are obtained without any problems later.)

We perform the summation with regard to $k$ and $n$. Then, the indices $k$ and $n$ in all the 
$(g_{\rm eff})^{tt}_{kn,\,lm}$, $(m_{\rm eff})_{kn,\,lm}$ and $\bar{\Delta}_{kn,\,lm}$ become  
$l$ and $m$ for the delta-functions in (\ref{ldcep}). Therefore, to shorten the expressions 
of equations, we in what follows denote these as 
\begin{eqnarray}
(g_{\rm eff})^{tt}_{lm,\,lm} \rightarrow  (g_{\rm eff})^{tt}_{lm}, \quad 
(m_{\rm eff})_{lm,\,lm} \rightarrow (m_{\rm eff})_{lm}, \quad 
\bar{\Delta}_{lm,\,lm} \rightarrow \bar{\Delta}_{lm}. 
\end{eqnarray}  

In what follows, $tt$- and $rr$-parts are basically same.  
We check $rr$-part only at the checkpoint. 
\newline

In (\ref{cagttpteg}), we consider to change the front factor $(2m)^2$ to $(2(m_{\rm eff})_{lm})^2$. 
For this we evaluate $\frac{(2m)^2}{(2(m_{\rm eff})_{lm})^2}\Lambda_{kn,\,lm}$. With 
\begin{eqnarray}
\frac{(2m)^2}{(2(m_{\rm eff})_{lm})^2} = 1- \frac{2\bar{\Delta}_{lm} }{m\,r}\varepsilon ^2+O(\varepsilon ^3),
\end{eqnarray} 
and $\Lambda_{kn,\,lm}$ given in (\ref{ldcep}), we can calculate in $\varepsilon^2$-order as 
\begin{subequations}\label{deftht}
\begin{align} 
\label{deftht0}
&\bullet \quad \frac{(2m)^2}{(2(m_{\rm eff})_{lm})^2}\Lambda_{lm}^{(0)}
=\Lambda_{lm}^{(0)}- \frac{2\bar{\Delta}_{lm} }{m\,r}\varepsilon ^2
+O(\varepsilon ^3) 
\equiv \Theta_{lm}^{(0)}, \\* 
\label{deflbd2}
&\bullet \quad \frac{(2m)^2}{(2(m_{\rm eff})_{lm})^2}\Lambda_{lm}^{(2)}
=\Lambda_{lm}^{(2)}
+O(\varepsilon ^3),\\* 
\label{deflbd4}
&\bullet \quad \frac{(2m)^2}{(2(m_{\rm eff})_{lm})^2}\Lambda_{lm}^{(4)}
=\Lambda_{lm}^{(4)}
+O(\varepsilon ^3). 
\end{align}
\end{subequations}
Therefore, we can write the $tt$-part in (\ref{cagttpteg}) as
\begin{subequations}
\label{dggdf}
\begin{align}
\label{dggdf4}
\textrm{(\ref{cagttpteg})} 
\,\,=&\hspace{6mm}
\sum_{l=0}^{l_{max}-4} \sum_{m=-l}^l
\int d^2x\, (2(m_{\rm eff})_{lm})^2 ( {\cal L}^{(0)}_{lm} +{\cal L}^{(2)}_{lm} +{\cal L}^{(4)}_{lm} )\\*
\label{dggdf2}
&+\sum_{l=l_{max}-3}^{l_{max}-2} \sum_{m=-l}^l
\int d^2x\, (2(m_{\rm eff})_{lm})^2  ( {\cal L}^{(0)}_{lm} +{\cal L}^{(2)}_{lm} )\\
\label{dggdf0}
&+\sum_{l=l_{max}-1}^{l_{max}} \sum_{m=-l}^l \int d^2x\, (2(m_{\rm eff})_{lm})^2 {\cal L}^{(0)}_{lm},\\*
\textrm{where} & \quad{\cal L}^{(K)}_{lm} 
\equiv
\Lambda_{lm}^{(K)} (g_{\rm eff})^{tt}_{l+K\,m} \partial_t \phi^*_{l+K\,m} \partial_t \phi_{lm} 
\quad \textrm{for $K=0,2,4$}, 
\end{align}
\end{subequations}
$l_{max}$ is finally taken to $\infty$. 
The $\phi_{lm}$ with $l$ larger than $l_{max}$ are zero, 
since no such $\phi_{lm}$ exist by definition. 
Calculation from (\ref{cagttpteg}) to (\ref{dggdf}) proceeds irrelevantly of either $tt$- or $rr$-part. 
\newline

Focusing on (\ref{dggdf4}), we write its integrand as
\begin{equation}
\label{fcco4}
\begin{split}
& 
\Omega_{lm}  \Theta_{lm}^{(0)} 
\bigg(
\partial_t \phi^*_{lm} \partial_t \phi_{lm}
+
\bigg(
\frac{(g_{\rm eff})^{tt}_{l+2m}}{(g_{\rm eff})^{tt}_{lm}}
\frac{\Lambda_{lm}^{(2)}}{\Theta_{lm}^{(0)}} 
\partial_t \phi^*_{l+2m}
+
\frac{(g_{\rm eff})^{tt}_{l+4m}}{(g_{\rm eff})^{tt}_{lm}}
\frac{\Lambda_{lm}^{(4)}}{\Theta_{lm}^{(0)}}
\partial_t \phi^*_{l+4m} 
\bigg)
\partial_t \phi_{lm} 
\bigg),
\\* 
& 
\Omega_{lm} \equiv (2(m_{\rm eff})_{lm})^2(g_{\rm eff})^{tt}_{lm}. 
\end{split}
\end{equation}
$\Theta_{lm}^{(0)}$ are in (\ref{deftht0}), and 
we defined $\Omega_{lm}$ to shorten the expression. 
Rescaling as
\begin{eqnarray}\label{cagtfdrd}
\phi_{lm} \to \frac{\phi_{lm} }{\big(\Theta_{lm}^{(0)}\big)^{1/2}}\quad \textrm{for all $l,\,m$}, 
\end{eqnarray} 
we can rewrite (\ref{fcco4}) as
\begin{eqnarray}\label{fcrsc4} 
\textrm{(\ref{fcco4})}
\!\!\! &=& \!\!\!  
\Omega_{lm} 
\bigg(
\partial_t \phi^*_{lm} \partial_t \phi_{lm}
+ 
\bigg(
\frac{(g_{\rm eff})^{tt}_{l+2\,m}}{(g_{\rm eff})^{tt}_{lm}}
\frac{\Lambda_{lm}^{(2)}}{\big(\Theta_{lm}^{(0)}\Theta_{l+2\,m}^{(0)}\big)^{1/2}}
\partial_t \phi^*_{l+2\,m}
\nonumber\\*
&&
\hspace{9mm}
+
\frac{(g_{\rm eff})^{tt}_{l+4\,m}}{(g_{\rm eff})^{tt}_{lm}}
\frac{\Lambda_{lm}^{(4)}}{\big(\Theta_{lm}^{(0)}\Theta_{l+4\,m}^{(0)}\big)^{1/2}}
\partial_t \phi^*_{l+4\,m}
\bigg)
\partial_t \phi_{lm}
\bigg).
\end{eqnarray}


Rescaling above can be understood as follows.
Writing $\phi_{lm}$ as $(\Theta_{lm}^{(0)})^{-1/2} (\Theta_{lm}^{(0)})^{1/2}\phi_{lm}$, 
treat as $\phi_{lm}^{(new)} \equiv (\Theta^{(0)}_{lm})^{1/2}\phi_{lm}$. 
At this time, $S[\phi]=S[(\Theta^{(0)})^{-1/2}\phi^{(new)}]$ 
but of course \linebreak ${\cal D}\phi \not= {\cal D}\phi^{(new)} = {\cal J}{\cal D}\phi$ formally writing, therefore, 
\begin{eqnarray}\label{lanc}
\int {\cal D}\phi \exp(-iS[\phi]) \not= \int {\cal D}\phi^{(new)} \exp(-iS[(\Theta^{(0)})^{-1/2}\phi^{(new)}]),   
\end{eqnarray} 
and ${\cal J}$ involves $r$ somehow as can be seen from (\ref{rrwjb}). 
However, ${\cal J}$ would be finally just a finite numerical value, although we could not evaluate its numerical value specifically. 
This is because our analysis is the one with the $\Delta r~(=r-r_{h,4D})$ assumed small, 
therefore the values of $r$ in our analysis are always some values slightly larger than $r_{h,4D}$.  
Therefore, in (\ref{lanc}), ${\cal J}$ is just some finite number and its r.h.s. can be finally written as
${\cal J}Z_{\phi}~(Z_{\phi}=\textrm{l.h.s. of (\ref{lanc})})$,
and the effect by the rescaling of  (\ref{cagtfdrd}) is irrelevant in the following analysis\footnote{ 
To understand this, let us consider the following path-integral for a limited section, $r_0 \le r \le r_1$, 
\begin{eqnarray}
&&
\mathcal{Z}_{\phi} \sim
\int {\cal D} \phi \exp (-i \int_{r_0}^{r_1} dr \, \phi(r) \,f(r) \,\phi(r)\, r^{-2}), \nonumber\\
&&
{\cal D}\phi \propto \prod_{\textrm{i=0 to 1}} d\phi (r_i), 
\quad 
\int_{r_0}^{r_1} dr \, \phi(r) \,f(r) \, \phi(r) \, r^{-2} \sim \sum_{\textrm{i=0 to 1}} \phi(r_i) \,f(r_i) \, \phi(r_i) \, r_i^{-2} \,\Delta r,\nonumber
\end{eqnarray} 
where $f(r)$ is some function with derivatives $\partial_r$. 
$r_i$ is the discretized coordinate, which represents some points on $r_0 \le r \le r_1$, and $\Delta r$ corresponds to $dr$. 
``$\sim$'' in the first line means, roughly saying, an expression having been reached performing path-integral for canonical momenta.  

Now, consider to change the variable of the path-integral as $\phi \to \phi^{(new)} \equiv \phi \, r^{-1}$. 
At this time, the expression of $\mathcal{Z}_{\phi}$ is changed as 
\begin{eqnarray}
&&  
\mathcal{Z}_{\phi}
\to
\prod_{\textrm{i=0 to 1}} r_i  \cdot \int {\cal D} \phi^{(new)} \exp (-i \int_{r_0}^{r_1} dr \, \phi^{(new)}(r) \,f(r) \,\phi^{(new)}(r)),
\nonumber \\
&& \!\!
\textrm{
where 
$ \displaystyle
\prod_{\textrm{i=0 to 1}} d\phi (r_i) \to \prod_{\textrm{i=0 to 1}} \,r_i\,d(\phi (r_i){r_i}^{-1})
=\prod_{\textrm{i=0 to 1}} \,r_i \cdot \prod_{\textrm{i=0 to 1}} \,d(\phi^{(new)} (r_i)).
$
}\nonumber
\end{eqnarray} 
$\prod_{\textrm{i=0 to 1}} \,r_i$ corresponds to ${\cal J}$, which is some finite number. 
The rescaling of (\ref{cagtfdrd}) is likewise.}. 
Even if ${\cal J}$ is a divergent quantity it is still a number, therefore the conclusion is not changed\footnote{ 
Let us just check the difference with the case of quantum anomalies.   
For any gauge transformation, $Z_{\phi}=Z_{\phi'}$.  
However, for some kinds of gauge transformation,
the path-integral measure is not invariant, 
which we formally represent as ${\cal D}\phi' = {\cal J}{\cal D}\phi$.
Action is also not invariant in some cases, which we represent as $S[\phi']=S[\phi]+\delta S[\phi]$.   
Therefore, $Z_{\phi'}=\int {\cal J}{\cal D} \phi \exp (-i (S[\phi]+\delta S[\phi]))=\int {\cal D} \phi \exp (\log {\cal J} -i (S[\phi]+\delta S[\phi]))$, 
from which $\log {\cal J} -i\delta S[\phi]=0$ is obtained.
In evaluating $\log {\cal J}$, some regularization is needed.  
As a result, $\log {\cal J}$ is given as some function of field strengths (e.g. around Eq.(5.20) in \cite{Fujikawa:2004cx}) 
and the quantum anomaly is given, which is the different point from the case of (\ref{cagtfdrd}).
}. In any case, following a general formula: $\int {\cal D} \phi \exp (-i \int dx^4\int dy^4\,\phi(x) {\cal M}(x,y)\phi(y)) \propto {\rm Det}\,{\cal M}$, 
there is no difference finally in the results of the path-integral obtained with/without (\ref{cagtfdrd}). 
\newline

It is important to give attention to the consideration above  that  $r$-direction is finite 
as it is related with the problem of boundary condition (this is the problem arising even in the classical level) and 
a fundamental supposition in the quantum field theory that spacetime spreads infinitely (breaking of translational 
symmetry at the boundary in $r$-direction may be needed to be cared, if to be exact).

What is being done in this section is just to obtain the expression of the action, 
not obtaining the solution, therefore these problems are irrelevant in this section. 
However, since we treat the quantum effect in the next section, we give attention to these problems.  

In the next section, we use the formulas of quantum anomaly. 
Although author has not checked the derivation process of these one by one entirely, 
these are obtained by once obtaining these in a flat spacetime, then by replacing the derivatives in these with covariant derivatives \cite{kimura}. 
In these derivation process, the supposition that the spacetime spreads infinitely would be used.  

Therefore, as what we do in fact, 
our study focuses on the vicinity of the horizon before taking the quantum effect, 
and quantum effect is taken in after focusing on the vicinity of the horizon, 
however, as the problem of how to consider, it would be possible to consider by the following way:  
{\bf 1)} performing the analysis for the quantum anomaly in an infinitely spreading flat spacetime, 
{\bf 2)} then replacing that flat spacetime with the one in this study with $r$-direction not limited, 
{\bf 3)} then just focusing on the vicinity of the horizon (not limiting $r$-direction),
{\bf 4)} we have used the formulas of quantum anomaly at there. 
By considering like this,  problems mentioned above are considered not to arise in the analysis in this study.
\newline

We here would like to look at the calculation 
from (\ref{fcco4}) to (\ref{fcrsc4}) via (\ref{cagtfdrd}) 
in the $rr$-part, since $\Theta_{lm}^{(0)}$ depend on $r$ as can be seen in (\ref{deftht0}), 
and at (\ref{fcrsc4}) in the calculation of the $rr$-part, the following equation appears, 
and which can be calculated as
\begin{eqnarray}\label{rrweptb}
&&\hspace{-1mm}
\frac{(g_{\rm eff})^{rr}_{l+K\,m}}{(g_{\rm eff})^{rr}_{lm}}
\frac{\Lambda_{lm}^{(K)}}{\Theta_{lm}^{(0)}}
\partial_r \Big(\frac{\phi_{l+K\,m}}{\big(\Theta_{l+K\,m}^{(0)}\big)^{1/2}}\Big)
\partial_r \Big(\frac{\phi_{lm}}{\big(\Theta_{lm}^{(0)}\big)^{1/2}}\Big)
\nonumber\\*
\!\! &&\hspace{-5mm} = 
\frac{(g_{\rm eff})^{rr}_{l+K\,m}}{(g_{\rm eff})^{rr}_{lm}}
\frac{\Lambda_{lm}^{(K)}}{\Theta_{lm}^{(0)}}
\Big(
\!
\frac{\partial_r\Theta_{l+K\,m}^{(0)}\phi_{l+K\,m}}{2\big(\Theta_{l+K\,m}^{(0)}\big)^{3/2}}
-\frac{\partial_r \phi_{l+K\,m}}{\big(\Theta_{l+K\,m}^{(0)}\big)^{1/2}}
\Big)  
\Big(
\!
\frac{\partial_r\Theta_{lm}^{(0)}\phi_{lm}}{2\big(\Theta_{lm}^{(0)}\big)^{3/2}}
-\frac{\partial_r \phi_{lm}}{\big(\Theta_{lm}^{(0)}\big)^{1/2}}
\Big)
\nonumber\\*
\!\! && \hspace{-5mm} = 
\frac{(g_{\rm eff})^{rr}_{l+K\,m}}{(g_{\rm eff})^{rr}_{lm}} 
\frac{\Lambda_{lm}^{(K)}}{\Theta_{lm}^{(0)}}
\frac{\partial_r \phi_{l+K\,m}}{\big(\Theta_{l+K\,m}^{(0)}\big)^{1/2}}
\frac{\partial_r\phi_{lm} }{\big(\Theta_{lm}^{(0)}\big)^{1/2}}+O(\varepsilon ^3)\,,
\end{eqnarray} 
where $K=2,4$ and
\begin{eqnarray}\label{rrwjb}
\frac{(g_{\rm eff})^{rr}_{l+K\,m}}{(g_{\rm eff})^{rr}_{lm}}\sim 1+\varepsilon ^2,\quad 
\Lambda_{lm}^{(K)} \sim \varepsilon^{K/2},\quad 
\Theta_{lm}^{(0)}\sim 1+\varepsilon+\left(1+r^{-1}\right)\varepsilon ^2, \quad 
\frac{\Lambda_{lm}^{(K)}}{\Theta_{lm}^{(0)}}\sim \varepsilon^{K/2},   
\end{eqnarray} 
from the definitions of (\ref{defgtt}), (\ref{ldcep}) and (\ref{deftht0}). 
Therefore, the extra terms drop and the $rr$-part at (\ref{fcrsc4}) can be obtained in the same way with (\ref{fcrsc4}) 
except $\partial_t$ and $(g_{\rm eff})^{tt}_{lm}$. 
\newline

Then, for the parts in (\ref{fcrsc4}),  
the following calculation can be held 
in  $\varepsilon^2$-order: 
\begin{eqnarray}\label{df2lbk} 
&&
\frac{(g_{\rm eff})^{tt}_{l+K\,m}}{(g_{\rm eff})^{tt}_{lm}} 
\frac{\Lambda_{lm}^{(K)}}{\big(\Theta_{lm}^{(0)}\Theta_{l+K\,m}^{(0)}\big)^{1/2}}
=\frac{\Lambda_{lm}^{(K)}}{\big(\Theta_{lm}^{(0)}\Theta_{l+K\,m}^{(0)}\big)^{1/2}}+O(\varepsilon ^3)
\equiv 2\overline{\Lambda}_{lm}^{(K)}
\\* 
&&\textrm{for all $l,\,m$, where $K=2,4$.}\nonumber  
\end{eqnarray} 
The one above can be actually checked with 
the $(g_{\rm eff})^{tt}_{lm}$, $\Lambda_{lm}^{(K)}$ and $\Theta_{lm}^{(0)}$ 
given in (\ref{defgtt}), (\ref{ldcep}) and (\ref{deftht0}) respectively, 
and can hold in the case of $rr$, 
namely if $(g_{\rm eff})^{tt}_{l+K\,m}$ and $(g_{\rm eff})^{tt}_{l\,m}$ 
are $(g_{\rm eff})^{rr}_{l+K\,m}$ and $(g_{\rm eff})^{rr}_{l\,m}$. 
Using (\ref{df2lbk}), we can write (\ref{fcrsc4}) as
\begin{eqnarray}\label{fcsgf4} 
(\ref{fcrsc4}) \,=\,
\Omega_{lm}
\big(
\partial_t \phi^*_{lm} \partial_t \phi_{lm}
+ 2
\big(
\overline{\Lambda}_{lm}^{(2)} \partial_t \phi^*_{l+2\,m} + \overline{\Lambda}_{lm}^{(4)} \partial_t \phi^*_{l+4\,m} 
\big)
\partial_t \phi_{lm}
\big). 
\end{eqnarray} 
We here define
\begin{eqnarray}
\Gamma_{lm}^{(2)} \equiv \overline{\Lambda}_{lm}^{(2)} \partial_t \phi_{l+2\,m}, 
\quad
\Gamma_{lm}^{(4)} \equiv \overline{\Lambda}_{lm}^{(2)} \partial_t \phi_{l+2\,m} + \overline{\Lambda}_{lm}^{(4)} \partial_t \phi_{l+4\,m},
\end{eqnarray}
to shorten the expression of the equations ($\Gamma_{lm}^{(2)}$ is not used immediately). 
Then, 
\begin{eqnarray}\label{fcfinl4}
(\ref{fcsgf4}) \,=\,
\Omega_{lm} 
\big(
\big|\partial_t \phi_{lm}+\Gamma_{lm}^{(4)}\big|^2
-\big(\overline{\Lambda}_{lm}^{(2)}\big)^2 \partial_t \phi^*_{l+2m}\partial_t \phi_{l+2m}
\big),
\end{eqnarray} 
where $\overline{\Lambda}_{lm}^{(K)}\sim\varepsilon^{K/2}$ from (\ref{df2lbk}) 
and
$
\Gamma_{lm}^{(4)}{}^*\Gamma_{lm}^{(4)} 
\,=\, \big(\overline{\Lambda}_{lm}^{(2)}\big)^2 \partial_t \phi^*_{l+2\,m}\partial_t \phi_{l+2\,m}+O(\varepsilon ^3)
$.\vspace{1.0mm}

Performing the calculation regarding (\ref{dggdf2}) likewise, we can write (\ref{dggdf}) as
\begin{eqnarray}\label{dgbjlh}
\textrm{(\ref{fcrsc4})} 
\! &=&\hspace{3.0mm} 
\sum_{l=0}^{l_{max}-4} \,\sum_{|m|=0}^l \int d^2x \,\Omega_{lm}
\big(
\big|\partial_t \phi_{lm}+\Gamma_{lm}^{(4)}\big|^2
-\big(\overline{\Lambda}_{lm}^{(2)}\big)^2 \partial_t \phi^*_{l+2m}\partial_t \phi_{l+2m}
\big)
\nonumber\\*
&&\!\!\!\!
+\sum_{l=l_{max}-3}^{l_{max}-2} \,\sum_{|m|=0}^l \int d^2x \,\Omega_{lm}
\big(
\big|\partial_t \phi_{lm}+\Gamma_{lm}^{(2)}\big|^2
-\big(\overline{\Lambda}_{lm}^{(2)}\big)^2 \partial_t \phi^*_{l+2m}\partial_t \phi_{l+2m}
\big)
\nonumber\\*
&&\!\!\!\!
+\sum_{l=l_{max}-1}^{l_{max}} \,\sum_{|m|=0}^l  \int d^2x \,\Omega_{lm}
\partial_t \phi_{lm}^* \partial_t \phi_{lm}+O(\varepsilon ^3). 
\end{eqnarray}
Calculation for the $rr$-part from (\ref{fcrsc4}) to (\ref{dgbjlh}) can be proceeded without problems, 
and the $rr$-part at (\ref{dgbjlh}) is also obtained basically same with (\ref{dgbjlh}). 
\newline

Now we consider to do uniformly slide each 
``$\Omega_{lm} \big(\overline{\Lambda}_{lm}^{(2)}\big)^2 \partial_t \phi^*_{l+2m} \partial_t \phi_{l+2m}$''  
appearing in the line of $l$ to the line of $l+2$ in (\ref{dgbjlh}). 
For this, let us check $\Omega_{lm}\big(\overline{\Lambda}_{lm}^{(2)}\big)^2$: 
\begin{eqnarray}
\Omega_{lm}\big(\overline{\Lambda}_{lm}^{(2)}\big)^2  
= \frac{405m^3({\cal I}^{A_2}_{lm})^2}{2\pi (r-2m)}{\varepsilon ^2}+O(\varepsilon ^3), 
\end{eqnarray}
where ${\cal I}^{A_2}_{lm}$ are numbers in (\ref{intp1}).
Thus we can write $\Omega_{lm}\big(\overline{\Lambda}_{lm}^{(2)}\big)^2$ changing its $l$ to $l+2$ as
\begin{eqnarray}\label{slpbl22}
\Omega_{lm}\big(\overline{\Lambda}_{lm}^{(2)}\big)^2  
= 
\frac{({\cal I}^{A_2}_{lm})^2}{({\cal I}^{A_2}_{l+2\,m})^2}
\Omega_{l+2\,m}
\big(\overline{\Lambda}_{l+2\,m}^{(2)}\big)^2
+O(\varepsilon ^3)
\equiv
\Omega_{l+2\,m} \Xi_{l+2\,m}, 
\end{eqnarray}
where
\begin{subequations}
\begin{align}
\frac{({\cal I}^{A_2}_{lm})^2}{({\cal I}^{A_2}_{l+2\,m})^2}
&=
\frac
{(2 l+9)(2 l+7)^2  \left((l+1)^2-m^2\right)\left((l+2)^2-m^2\right)}
{(2 l+1) (2 l+3)^2\left((l+3)^2-m^2\right)\left((l+4)^2-m^2\right)},
\\*
\Xi_{l+2\,m} 
&\equiv
\frac{({\cal I}^{A_2}_{lm})^2}{({\cal I}^{A_2}_{l+2\,m})^2}\big(\overline{\Lambda}_{l+2\,m}^{(2)}\big)^2.
\end{align} 
\end{subequations}
With (\ref{slpbl22}), we can replace as
\begin{eqnarray}
\Omega_{lm}
\big(\overline{\Lambda}_{lm}^{(2)}\big)^2 
\partial_t \phi^*_{l+2\,m}\partial_t \phi_{l+2\,m} 
\rightarrow 
\Omega_{l+2\,m} \Xi_{l+2\,m} \partial_t \phi^*_{l+2\,m}\partial_t \phi_{l+2\,m} \quad \textrm{for all $l$, $m$}.
\end{eqnarray}
Therefore, uniformly sliding each  
``$\Omega_{lm} \big(\overline{\Lambda}_{lm}^{(2)}\big)^2 
\partial_t \phi^*_{l+2m} \partial_t \phi_{l+2m}$'' by $2$ regarding $l$ in (\ref{dgbjlh}),  
\begin{eqnarray}\label{erslh}
\textrm{(\ref{dgbjlh})} 
\!\! &=&
\hspace{6.0mm} 
\sum_{l=0}^{1} \sum_{|m|=0}^l
\int d^2x 
\,\Omega_{lm}
\big|\partial_t \phi_{lm}+\Gamma_{lm}^{(4)}\big|^2
\nonumber\\*[3.0mm]
&&
+ \,
\sum_{l=2}^{l_{max}-4} 
\sum_{|m|=0}^{l-2} 
\int d^2x 
\,\Omega_{lm}
\big(
\big|\partial_t \phi_{lm}+\Gamma_{lm}^{(4)}\big|^2
-\Xi_{lm} \, \partial_t \phi^*_{lm}\partial_t \phi_{lm}
\big)
\nonumber\\*[-1.5mm]
&& 
+ \, 
\sum_{l=2}^{l_{max}-4} 
\sum_{|m|=l-1}^l
\int d^2x  
\,\Omega_{lm}
\big|\partial_t \phi_{lm}+\Gamma_{lm}^{(4)}\big|^2
\nonumber\\*[3.0mm]
&&
+ \, 
\sum_{l=l_{max}-3}^{l_{max}-2} \sum_{|m|=0}^{l-2}
\int d^2x 
\,\Omega_{lm}
\big(
\big|\partial_t \phi_{lm}+\Gamma_{lm}^{(2)}\big|^2
-\Xi_{lm} \, \partial_t \phi^*_{lm}\partial_t \phi_{lm}
\big)
\nonumber\\*[-1.5mm]
&&
+ \,
\sum_{l=l_{max}-3}^{l_{max}-2}  \sum_{|m|=l-1}^l
\int d^2x 
\,\Omega_{lm}
\big|\partial_t \phi_{lm}+\Gamma_{lm}^{(2)}\big|^2
\nonumber\\[3.0mm]
&&
+ \,
\sum_{l=l_{max}-1}^{l_{max}} \sum_{|m|=0}^{l-2} 
\int d^2x 
\,\Omega_{lm}
\big(
\partial_t \phi_{lm}^* \partial_t \phi_{lm}     
-\Xi_{lm} \, \partial_t \phi^*_{lm}\partial_t \phi_{lm}
\big) 
\nonumber\\*[-1.5mm]
&&
+ \,
\sum_{l=l_{max}-1}^{l_{max}} \sum_{|m|=l-1}^l
\int d^2x 
\,\Omega_{lm}
\partial_t \phi_{lm}^* \partial_t \phi_{lm}+O(\varepsilon ^3).
\end{eqnarray}
We once again perform the rescaling of the fields as
\begin{eqnarray}\label{ppfdtpg}
&&\phi_{l\,m}\rightarrow 
\frac{\phi_{l\,m}}
{(1-\Xi_{lm})^{1/2}} 
\quad
\textrm{for $l=2,3,\cdots,l_{max}$ ($l=0,1$ are not included)} 
\nonumber\\*[-4mm]
&&\hspace{38.0mm}\textrm{and $|m|=0,1,\cdots,l-2$ for each $l$.} 
\end{eqnarray}
This rescaling is possible by the same reason written around (\ref{lanc}). 
At this time,  $\overline{\Lambda}_{lm}^{(K)}\sim\varepsilon^{K/2}$ (see under (\ref{fcfinl4})) 
and $\Gamma_{lm}^{(K)}$ and $\partial_t \phi\,\Gamma_{lm}^{(K)*}$ can stay same in  $\varepsilon^2$-order as
\begin{eqnarray}\label{ppfdtgm}
&&\Gamma_{lm}^{(K)} \rightarrow \Gamma_{lm}^{(K)}+O(\varepsilon ^3),  
\quad 
\partial_t \phi\,\Gamma_{lm}^{(K)*} 
\rightarrow 
\partial_t \phi\,\Gamma_{lm}^{(K)*}+O(\varepsilon ^3) \\*[2mm]
&&\textrm{for all $l$ and $m$, where $K=2,4$.}
\nonumber
\end{eqnarray}
Therefore, we can exchange the lines in (\ref{erslh}) with the squared form as
\begin{eqnarray}\label{ikud}
\big|\partial_t \phi_{lm}+\Gamma_{lm}^{(K)}\big|^2
-\Xi_{lm} \, \partial_t \phi^*_{lm}\partial_t \phi_{lm}
\rightarrow
\big|\partial_t \phi_{lm}+\Gamma_{lm}^{(K)}\big|^2,
\end{eqnarray} 
where ``$\rightarrow$'' means the rescaling (\ref{ppfdtpg}). 
Therefore, we can write (\ref{erslh}) as
\begin{eqnarray}\label{esqcml} 
\textrm{(\ref{erslh})} 
\!\! 
&=&
\hspace{6mm}
\sum_{l=0}^{l_{max}-4} \sum_{m=-l}^l 
\int d^2x \,\Omega_{lm} 
\big|\partial_t \phi_{lm}+\Gamma_{lm}^{(4)}\big|^2
\nonumber\\*
&&
+\, 
\sum_{l=l_{max}-3}^{l_{max}-2} \sum_{m=-l}^l
\int d^2x \,\Omega_{lm} 
\big|\partial_t \phi_{lm}+\Gamma_{lm}^{(2)}\big|^2
\nonumber\\*
&&
+\,
\sum_{l=l_{max}-1}^{l_{max}} \sum_{m=-l}^l 
\int d^2x \,\Omega_{lm} \partial_t \phi_{lm}^* \partial_t \phi_{lm}.      
\end{eqnarray}

We here would like to give attention to the $rr$-part. 
The point to be checked between (\ref{dgbjlh}) and (\ref{esqcml}) is the manipulation (\ref{ppfdtpg}): whether 
$
\partial_r \big( \frac{\phi_{l\,m}}{(1-\Xi_{lm})^{1/2}} \big) = 
\frac{\partial_r \phi_{l\,m}}{(1-\Xi_{lm})^{1/2}} 
+O\big(\varepsilon ^3\big)
$ 
can be held or not at (\ref{ppfdtpg}) in the calculation of the $rr$-part.  
For this, let us check the $r$-dependence of  $\Xi_{lm}$:
$\Xi_{lm}
= 
\frac{({\cal I}^{A_2}_{lm})^2}{({\cal I}^{A_2}_{l+2\,m})^2}
\Big(
\frac
{\Lambda_{lm}^{(2)}}
{2\big(\Theta_{lm}^{(0)}\Theta_{l+2\,m}^{(0)}\big)^{1/2}}
\Big)^2$,
where $\frac{({\cal I}^{A_2}_{lm})^2}{({\cal I}^{A_2}_{l+2\,m})^2}$ are numbers, 
$\Lambda_{lm}^{(2)}\sim \varepsilon$ and \linebreak $\Theta_{lm}^{(0)}\sim  1+\varepsilon+(1+r^{-1}) \varepsilon ^2$. 
Therefore, $\Xi_{lm}$ is independent of $r$ in $\varepsilon^2$-order. 
If so, the equation above can be held, and the $rr$-part at (\ref{esqcml}) can be obtained as (\ref{esqcml}) 
as well as the $tt$-part without the difference of $\partial_r$ and  $(g_{\rm eff})^{rr}_{lm}$.  

Since it can be written as follows:
\begin{subequations}
\begin{align}
\partial_t \phi_{lm}+\Gamma_{lm}^{(4)} \,=&\,\,
\partial_t(\phi_{lm}+\overline{\Lambda}_{lm}^{(2)}\phi_{l+2\,m}+\overline{\Lambda}_{lm}^{(4)}\phi_{l+4\,m}), \\ 
\partial_t \phi_{lm}+\Gamma_{lm}^{(2)} \,=&\,\,
\partial_t(\phi_{lm}+\overline{\Lambda}_{lm}^{(2)}\phi_{l+2\,m}),
\end{align}
\end{subequations} 
let us perform the redefinition of the fields as  
\begin{subequations}\label{rcmbmt} 
\begin{align}
&\bullet 
\quad 
\varphi_{lm} \equiv \,
\phi_{lm}+\overline{\Lambda}_{lm}^{(2)}\phi_{l+2\,m}+\overline{\Lambda}_{lm}^{(4)}\phi_{l+4\,m}  
\quad 
\textrm{for $l=0,1,\cdots,l_{max}-4$,}
\\* 
&\bullet 
\quad   
\varphi_{lm} \equiv \,
\phi_{lm}+\overline{\Lambda}_{lm}^{(2)}\phi_{l+2\,m}   
\qquad\qquad\qquad\hspace{2mm} \,
\textrm{for $l=l_{max}-3,l_{max}-2$,}
\\* 
&\bullet \quad 
\varphi_{lm} \equiv \,
\phi_{lm}
\qquad\qquad\qquad\qquad\qquad\qquad \,\,
\textrm{for $l=l_{max}-1,l_{max}$,}
\end{align}
\end{subequations} 
where $m$ above are $0,\pm 1,\cdots,\pm(l-2)$ for each $l$. 
The leadings of $\overline{\Lambda}_{lm}^{(K)}$ is $\varepsilon^{K/2}$.
 
(\ref{cagtfdrd}) and (\ref{ppfdtpg}) are rescalings 
which can be absorbed as configurations of the path-integral for $\phi_{lm}$ for the reason written around (\ref{lanc}), however (\ref{rcmbmt}) is recombinations.  
Therefore, the Jacobian for $\phi_{lm} \rightarrow \varphi_{lm}$, 
should be checked. 
Forming a matrix according to (\ref{rcmbmt})
We can check it gives  unit.

With $\varphi_{lm}$ above, we can finally obtain the decoupled 2D effective action which is 
equivalent with (\ref{cncwra}) as a action in the range of $\varepsilon ^2$ as
\begin{eqnarray}\label{ef2daccp}
(\ref{esqcml}) \, =  \,
\sum_{l=0}^{l_{max}} \sum_{|m|=0}^l 
\int d^2x \, \Phi_{lm}
(  
(g_{\rm eff})^{tt}_{lm} \, \partial_t \varphi^*_{lm}  \partial_t  \varphi_{lm}
+(g_{\rm eff})^{rr}_{lm} \, \partial_r  \varphi^*_{lm}  \partial_r  \varphi_{lm}
),
\end{eqnarray}
where 
{\bf 1)} $\Phi_{lm} = \big(2(m_{\rm eff})_{lm}\big)^2$ and the 2D effective metrices are given in (\ref{defgmtpa}).  
{\bf 2)} Einstein equation can be satisfied with these effective metrices. 
{\bf 3)} Since the labels distinguishing the effective metrices are irrelevant of the spins, 
the effective metrices would not be changed if we considered fermions \cite{Li:2010eca,Becar:2011fc,Mao:2011zzb} 
and higher spin fields \cite{Iso:2007nf}. 

Lastly, the behavior (\ref{rrwjb}) is critical in the feasibility of the analysis in this subsection.

\section{Hawking Temperature in the effective 2D}  
\label{sec:hwkttd}
 
We have obtained the 2D effective metrices, 
which are labeled by spherical harmonics modes.  
From these, we can naively expect 
{\bf 1)} existence of various Hawking temperatures for each effective metric,  
{\bf 2)} correspondingly, breaking of the zeroth law of the black hole thermodynamics.
(Furthermore, 
{\bf 3)}  $(m_{\rm eff})_{kn,\,lm}$ get depended on $r$ as in $(\ref{mdfeft})$, though this is not problems in the analysis's order.) 
Hence, let us check the Hawking temperature.

We can obtain the position of the horizon in the 2D  picture from $(g_{\rm eff})_{{tt},\,{lm}}=0$ as
\begin{eqnarray}\label{peeh}
(r_{h,2D})_{lm}=2m+\frac{15  m \,  {\cal I}^{C_0}_{lm}}{8 \pi }\,\varepsilon ^2+O(\varepsilon ^3).
\end{eqnarray} 
As this is labeled by spherical harmonics modes, we can expect the points above. 
However, the Hawking temperature  obtained from the 2D effective metrices 
with the one above is
\begin{eqnarray}\label{srfsg}
T_H = {1}/{8\pi m}+O(\varepsilon ^3).
\end{eqnarray} 
This is just that in the 4D Schwarzschild 
and free from the concerns above.
The reason of this is the same with those written in Sec.\ref{subsec:hwkkdk}, 
where it is considered replacing with (\ref{sdghc}).
 
The original 4D and effective 2D spacetimes are different each other.  
However the Hawking temperature in the effective 2D spacetime 
is generally considered to coincide with the one in the original 4D spacetime.  
Actually the one above coincides with (\ref{hkod4d}).

\section{Hawking flux by anomaly cancellation}
\label{sec:hfac}

We call the anomaly cancellation method as ``anomaly cancellation''.  
Since our U(1) gauge field  does not arise  chiral anomalies, 
we ignore it as mentioned under (\ref{kkew}). 
Hence, we do not consider the Hawking flux of the electric charged current.

\subsection{Set up of the radial direction}
\label{subsec:seprad}
 
The key point in the anomaly cancellation is the fact of no outgoing modes 
on the horizon at the classical level. To treat this situation in the anomaly 
cancellation, some interval from $(r_{h,2D})_{lm}$ in the radial direction are sharply divided as follows:\footnote{ 
Radial direction is sharply divided with $\epsilon$ in all the papers of the anomaly cancellation, which is unnatural.
This problem is commented in Chap.4 in \cite{Umetsu:2010ts} and treated in \cite{Umetsu:2008cm}.  
There is one more artificial point in the anomaly cancellation, 
which is to use two anomalies,  gravitational and consistent anomalies.  
\cite{Banerjee:2007uc} cares this point. 
}
\begin{subequations}\label{rgn} 
\begin{align}
\label{rgnh} (r_{h,2D})_{lm}           \leq  & \, r  \leq (r_{h,2D})_{lm} + \epsilon_{lm}, \\* 
\label{rgno} (r_{h,2D})_{lm}+ \epsilon_{lm} <  & \,  r \leq (r_o)_{lm}.           
\end{align}
\end{subequations}
\begin{itemize}
\item $\epsilon_{lm}$ represent the divided points, which are finally taken to zero,
\item $(r_o)_{lm}$ mean the positions put by hand reasonably supposing that it is the maximum of the $r$ 
to where the description by the 2D effective action (\ref{ef2daccp}) is possible. 
\item (\ref{rgnh}) is the region where supposed only ingoing modes exist at the classical level, 
\item (\ref{rgno}) is the region where both ingoing and outgoing modes exist at  classical level. 
\end{itemize}

We refer to the two regions, (\ref{rgnh}) and (\ref{rgno}), as the regions $\cal H$ and $\cal O$, respectively. 
In what follows we suppose the following corresponding in the 2D effective picture: 
\begin{eqnarray} 
&&\textrm{the outgoing modes $\to$ the right-hand modes,} \nonumber\\* 
&&\textrm{the ingoing modes $\to$ the left-hand modes.} \nonumber
\end{eqnarray}

\subsection{Hawking flux of the energy-momentum tensors} 
\label{subsec:femt}

We consider the distribution function
in the region $(r_{h,2D})_{lm}  \leq  r  \leq (r_{o})_{lm}$ as
\begin{eqnarray}\label{dsfgef}
Z\left[ (g_{\rm eff})_{lm}^{\mu\nu},\Phi_{lm} \right]
=\int{\cal D} \varphi_{lm} \, \exp i S_{\textrm{2D}} ( (g_{\rm eff})_{lm}^{\mu\nu},\Phi_{lm},\varphi_{lm} ),
\end{eqnarray}
where $(g_{\rm eff})_{lm}^{\mu\nu}$, $\Phi_{lm}$ and $\varphi_{lm}$ are those in (\ref{ef2daccp}).  
Then, consider degrees of gauge freedom of general coordinate transformation in that region as 
\begin{eqnarray}\label{gdtsf}
x^\mu \mapsto x'{}^\mu = x^\mu - \eta^\mu\left(x^\mu\right). 
\end{eqnarray}

Variation toward these can be written as
\begin{eqnarray}\label{coordid}
(\delta  Z)_{lm} =
\Big( 
\delta_L  (g_{\rm eff})_{lm}^{\mu\nu} \frac{\delta}{\delta_L  (g_{\rm eff})_{lm}^{\mu\nu}} 
+ \delta_L  A_{\mu\, lm} \frac{\delta}{\delta_L  A_{\mu,\, lm}} 
+\delta_L  \Phi_{lm} \frac{\delta}{\delta_L  \Phi_{lm}}  \Big) Z   
\end{eqnarray} \vspace{-4mm}
\begin{subequations}  
\begin{align}
\textrm{with} \quad 
\label{ldvgmn}
\delta_L (g_{\rm eff})_{lm}^{\mu\nu} 
= & \, 
-( \nabla^\mu_{lm} \eta^\nu + \nabla^\nu_{lm} \eta^\mu_{lm}), 
\\* 
\label{ldvamn}
 \delta_L  A_{\mu,\, lm} 
= & \, 
\nabla_{\mu,\, lm} \eta^\nu,\, A_{\nu \, lm} 
+ \eta^\nu \nabla_{\nu,\, lm} A_{\mu,\, lm}, 
\\* 
( 
\delta_L A^\mu_{lm}  = 
& \, 
-\nabla^\mu_{lm}  
\eta^\nu,\, A_{\nu \, lm}+ \eta^\nu \nabla_{\nu,\, lm} A^\mu_{lm} )
\nonumber 
\\* 
\label{ldvpnn}
\delta_L  \Phi_{lm}=& \, \, \eta^\mu \partial_\mu \Phi_{lm},  
\end{align}
\end{subequations} 
where $\delta_L$ means Lie derivative. 
$l$ and $m$ are not summed.  
We keep $A_{\mu,\, lm}$ just in case. 

Each $(\delta  Z)_{lm}$ should vanishes, from which we can obtain the conservation laws for the energy-momentum tensors at the classical level from $(\delta  Z)_{lm} =0$. 
Aside from these, quantum anomalies exist as \cite{Bertlmann:2000da}
\begin{subequations}  \label{gran}
\begin{align}    
\label{gran1}  
\nabla_\mu T^\mu{}_{\nu,\, lm}  
&= 
\pm \frac{1}{96\pi (-(g_{\rm eff})_{lm})^{1/2}}\epsilon^{\beta\delta} \partial_\delta \partial_\alpha \Gamma^\alpha_{\nu \beta,\,lm}
\equiv \mathscr{A}_{\nu,\,lm}^{\pm}, 
\\*  
&
\hspace{-13mm}\textrm{($+/-$ $\to$ left- / right-hand mode's contributions)} 
\nonumber \\*
\label{gran2}
\nabla_\mu \widetilde{T}^\mu{}_{\nu,\, lm}   
&= \mp \frac{1}{96\pi (-(g_{\rm eff})_{lm})^{1/2}}\epsilon_{\mu\nu}\partial^\mu R_{lm}
\equiv \widetilde{\mathscr{A}}_{\nu,\,lm}^{\mp}, 
\\*    
&\hspace{-13mm}\textrm{($-/+$ $\to$ left- / right-hand mode's contributions)} 
\nonumber 
\end{align}
\end{subequations}
where $\epsilon^{tr}=1$ and $\epsilon_{\mu\nu}=(g_{\rm eff})_{\mu\alpha,\,lm} (g_{\rm eff})_{\nu\beta,\, lm} \epsilon^{\alpha\beta}$. 
Top and bottom are the consistent and covariant anomalies. 
$\widetilde{T}^\mu{}_{\nu,\, lm}$ follow the boundary condition as 
\begin{eqnarray}\label{bctwt} 
(\widetilde{T}_H)_{\mu \nu,\,lm} \big|_{r=(r_{h,2D})_{lm}}=0. 
\end{eqnarray}
The conservation laws in the anomaly cancellation are given combining these as
\begin{subequations}\label{lfybks}
\begin{align}   
\label{lfybks1}
\nabla_\mu T^\mu{}_{\nu,\, lm} =& \,
F_{\mu\nu,\, lm}J^\mu_{lm} +A_{\nu,\,lm} \nabla_\mu J^\mu_{lm}
-\frac{\partial_\nu \Phi_{lm}}{(-(g_{\rm eff})_{lm})^{1/2}}\frac{\delta S_{\textrm{2D}}}{\delta_L \Phi_{lm}}
+ \textrm{both/either $\mathscr{A}_{\nu,\, lm}^{\pm}$}, 
\\*
\label{lfybks2}
\nabla_\mu \widetilde{T}^\mu{}_{\nu,\, lm} =& \,
 F_{\mu\nu,\, lm}J^\mu_{lm} +A_{\nu,\,lm} \nabla_\mu J^\mu_{lm}
-\frac{\partial_\nu \Phi_{lm}}{(-(g_{\rm eff})_{lm})^{1/2}}\frac{\delta S_{\textrm{2D}}}{\delta_L \Phi_{lm}}
+ \textrm{both/either $\widetilde{\mathscr{A}}_{\nu,\, lm}^{\mp}$}, 
\end{align} 
\end{subequations} 
where  
$J^\mu_{lm} \equiv \frac{1}{(-(g_{\rm eff})_{lm})^{1/2}}\frac{\delta S_{\textrm{2D}}}{\delta_L A_{\mu,\,lm}}$ and 
$T_{\mu\nu,\,lm} \equiv \frac{2}{(-(g_{\rm eff})_{lm})^{1/2}}\frac{\delta S_{\textrm{2D}}}{\delta_L (g_{\rm eff})^{\mu\nu}_{lm}}$. 
``both'' or ``either'' is taken according to both  left- and right-hand modes exist or not.  
Anomalies vanish in ``both'' 
as the left- and right-hand modes  cancel  each other. 

We show (\ref{lfybks}) in our case by calculating these for the case $\nu=t$ and $r$ respectively 
using (\ref{defgtt}), (\ref{defgrr}) and (\ref{defgtt}) etc as  
\begin{subequations} \label{clssan}
\begin{align}
\label{clssan1}
&\partial_r T^r{}_{t,\,lm} =\textrm{both/either $\mathscr{A}_{t,\, lm}^{\pm}$} (=\pm\partial_r N^r{}_{t,\,lm}),
\quad 
\partial_r T^r{}_{r,\,lm} = 0, 
\\* 
\label{clssan2}
&\partial_r \widetilde{T}^\mu{}_{t,\,lm} = \textrm{both/either $\widetilde{\mathscr{A}}_{t,\, lm}^{\mp}$} (=\pm\partial_r \widetilde{N}^r{}_{t,\,lm}),
\quad  
\partial_r \widetilde{T}^r{}_{r,\,lm} = 0,\\* 
&N^r{}_{t,\,lm} = (f'^2+f f'')/192\pi, \quad 
\widetilde{N}^r{}_{t,\,lm} = (f f''-(f')^2/2)/96\pi, 
\nonumber
\end{align} 
\end{subequations} 
where $f$ means $-(g_{\rm eff})_{tt,\,lm}$ and $'$ means $\partial_r$. 
We have used the facts that our gauge fields are ignoble (see under (\ref{kkew}))  
and our dilaton is time-independent with our killing vector.

We give the expressions of the energy-momentum tensors we employ as  
\begin{subequations} 
\label{tmal}
\begin{align}
\label{tmu1}
T^\mu{}_{\nu,\, lm} =& \, (T_o)^\mu{}_{\nu,\, lm} \Theta_{lm} + (T_H)^\mu{}_{\nu,\, lm} H_{lm},
\\* 
\label{tmu2}
\widetilde{T}^\mu{}_{\nu,\, lm} =& \, (\widetilde{T}_o)^\mu{}_{\nu,\, lm} \Theta_{lm} + (\widetilde{T}_H)^\mu{}_{\nu,\, lm} H_{lm},
\end{align}
\end{subequations} 
where $\Theta_{lm}$ mean the step function $ \theta \left(r- \left((r_{h,2D})_{lm}  + \epsilon_{lm} \right)\right)$ and 
$H_{lm}$ is $1-\Theta_{lm}$. 
Therefore, 
\begin{subequations} 
\label{factemt}
\begin{align}
&\textrm{$(T_H)^\mu{}_{\nu,\, lm}$ include only the right-hand modes},\\* 
&\textrm{$(T_o)^\mu{}_{\nu,\, lm}$ include both hand modes, which leads no anomalies.}  
\end{align}
\end{subequations}  
$(\widetilde{T}_{H})^\mu{}_{\nu,\, lm}$ and $(\widetilde{T}_{o})^\mu{}_{\nu,\, lm}$ are likewise. 
Sharp expressions of  (\ref{tmal})  is rooted in setting (\ref{rgn}).

From (\ref{clssan}) and (\ref{tmal}) with (\ref{factemt}), we can obtain the identities we consider as 
\begin{subequations}  \label{gwiblf}
\begin{align}
\label{gwiblf1}
&\partial_r  (T_H)^r{}_{t,\, lm} = \,  \mathscr{A}_{t,\,lm}^{+}=\partial_r N^r{}_{t,\,lm} 
\quad {\rm and} \quad  
\partial_r  (T_o)^r{}_{t,\, lm} = \, 0, \\* 
\label{gwiblf2}
&\partial_r  (\widetilde{T}_H)^r{}_{t,\, lm} = \,  \widetilde{\mathscr{A}}_{t,\,lm}^{-}-\partial_r \widetilde{N}^r{}_{t,\,lm}
\quad {\rm and} \quad 
\partial_r  (\widetilde{T}_o)^r{}_{t,\, lm} = \, 0.
\end{align}
\end{subequations}

From these, we can get the expressions of 
$(T_{H,\,o})^r{}_{t,\, lm}$ and 
$(\widetilde{T}_{H,\,o})^r{}_{t,\, lm}$ as 
\begin{subequations}   
\label{cogran} 
\begin{align} 
\label{cogran1} 
& (T_H)^r{}_{t,\, lm} = (c_H)^r{}_{t,\, lm} 
+ \int_{(r_{h,2D})_{lm}}^r \!\!\!\! dr\,\partial_r N^r{}_{t,\,lm}, 
\quad (T_o)^r{}_{t,\, lm} = (c_o)^r{}_{t,\, lm},\\*
\label{fogran1} 
& (\widetilde{T}_H)^r{}_{t,\, lm} = (\widetilde{c}_H)^r{}_{t,\, lm} 
+ \int_{(r_{h,2D})_{lm}}^r \!\!\!\! dr\,\partial_r\widetilde{N}^r{}_{t,\,lm}, 
\quad
(\widetilde{T}_o)^r{}_{t,\, lm} = (\widetilde{c}_o)^r{}_{t,\, lm}, 
\end{align}
\end{subequations}  
where 
($(c_H)^r{}_{t,\, lm}$, $(\widetilde{c}_H)^r{}_{t,\, lm}$) and  
($(\widetilde{c}_o)^r{}_{t,\, lm}$, $(\widetilde{c}_o)^r{}_{t,\, lm}$) are integral constants.  
The former two are the values of those at horizon, 
the latter two are the values of those at $r=(r_o)_{lm}$. 
$(\widetilde{c}_o)^r{}_{t,\, lm}$ is identified with the total amount of the Hawking flux 
(e.g. \cite{Iso:2006wa,Iso:2006ut}\footnote{
$(\widetilde{c}_o)^r{}_{t,\, lm}$ can be identified 
with the value of the total amount of the black-body radiation 
through the identification of $T^r{}_t$ with that ((28) in \cite{Iso:2006wa}), 
where the fermion case is considered in \cite{Iso:2006wa,Iso:2006ut} 
to avoid the problem of superradiance supposing that it would be the 
same with the bosonic case.  

Once one has checked that 
the value of $T^r{}_t$ can agree to the black-body radiations 
in the 3 kinds of the fundamental 4D black holes (Schwarzschild, Kerr and charged), 
all the papers concerning the anomaly cancellation compute 
the value of $T^r{}_t$ in various black holes, 
and consider that it always represents the total amount of the black-body radiation. 
We in this study also follow this way. 
}).

We consider an equation obtained from (\ref{cogran}) as\footnote{
There is one point. We can see a quantity: 
$(\widetilde{T}_H)^r{}_{t,\, lm}  -\left( (\widetilde{c}_o)^r{}_{t,\, lm} - (c_o)^r{}_{t,\, lm} \right)$ 
appears when obtaining (\ref{enrb}) from (\ref{cogran}). 
We can see we should redefine it as a new 
$(\widetilde{T}_H)^r{}_{t,\, lm}$ so that new $(\widetilde{T}_H)^r{}_{t,\, lm}$ can vanish at the horizon 
as in (\ref{bctwt}) by appropriately taking the integral constants, $(\widetilde{c}_o)^r{}_{t,\, lm}$ and $(c_o)^r{}_{t,\, lm}$. 

This is because $(\widetilde{T}_H)^r{}_{t,\, lm}$ should vanish at the horizon to get (\ref{enrb})
(or (24) in \cite{Iso:2006wa} or (36) in \cite{Iso:2006ut}), however it does not if it is as it is. 
We can see this as
$(\widetilde{T}_H)^r{}_{t,\, lm} \big|_{r=(r_{h,2D})_{lm}} = - \widetilde{N}^r{}_{t,\,lm} \big|_{r=(r_{h,2D})_{lm}} \not=0$ with $f'\big|_{r=(r_{h,2D})_{lm}}\not=0$. 
(This is not written in any papers such as \cite{Iso:2006wa,Iso:2006ut}. 
Further, \cite{Bertlmann:2000da} is referred at (36) in \cite{Iso:2006ut}, 
so look it. Then its (6.21) corresponds to (\ref{enrb}). 
There should be some integral constants there when ${\cal P}_{\mu\nu}$ is obtained by performing integration in (6.22), 
however  no comment about this point there.) 
}
\begin{eqnarray}\label{enrb}
(\widetilde{T}_H)^r{}_{t,\, lm} - (T_H)^r{}_{t,\, lm} =(f f''-2(f')^2)/192\pi. 
\end{eqnarray}    
We can obtain the value of $(c_H)^r{}_{t,\, lm}$ from (\ref{enrb}) with (\ref{bctwt}) as
\begin{eqnarray}\label{wcdah} 
(c_H)^r{}_{t,\, lm}   
= -\frac{1}{192\pi}(f f''-2(f')^2)\big|_{r=(r_{h,2D})_{lm}}  
= \frac{(f')^2}{96\pi}\Big|_{r=(r_{h,2D})_{lm}}
= \frac{\pi}{6}T_H^2,
\end{eqnarray} 
where $f'|_{r=(r_{h,2D})_{lm}}=4\pi T_H$, 
($T_H$ is 
(\ref{srfsg})).  
Variation 
for (\ref{gdtsf}) can be written as 
\begin{eqnarray}\label{tkskt}
(\delta S_{\textrm{2D}})_{lm} 
\!\!\! &=& \!\!\! -\int d^2x (-(g_{\rm eff})_{lm})^{1/2}\,\eta^\nu \nabla_{\mu,\,lm} T^\mu{}_{\nu,\,lm}\nonumber\\*
\!\!\! &=& \!\!\! -\int dx^2 \, \eta^t 
( 
\hspace{2.5mm}
( (T_o)^r{}_{t,\, lm} - (T_H)^r{}_{t,\, lm}  + N^r{}_{t,\,lm}) 
\delta \left( r-((r_{h,2D})_{lm} + \epsilon_{lm}) \right)
\nonumber\\ 
&&\qquad \qquad\quad \! +\partial_r (N^r{}_{t,\,lm} H)
). 
\end{eqnarray} 
$\epsilon_{lm}$ are taken to zero as the near-horizon limit.  
The last term will vanish \cite{Robinson:2005pd,Iso:2006wa,Murata:2006pt}\footnote{  
$\displaystyle \int_{(r_{h,2D})_{lm}}^{(r_{h,2D})_{lm}+\epsilon} dr \, \partial_r \, (\cdots) \to 0$ with $\epsilon \to 0$.
}.
  
$(\delta S_{\textrm{2D}})_{lm}$ should vanish,  
from which $(c_o)^r{}_{t,\, lm}$, the total amount of the Hawking flux, are determined as
\begin{eqnarray}\label{ea0ahmn}
(c_o)^r{}_{t,\, lm} 
= (c_H)^r{}_{t,\, lm}  -  N^r{}_{t,\,lm} \big|_{r=(r_{h,2D})_{lm}}
= \pi T_H^2/12. 
\end{eqnarray}

This result is the same with just the Schwarzschild \cite{Robinson:2005pd,Iso:2006wa,Iso:2006ut}.   
The reason of this is that 
the Hawking flux is determined from the $f$, $f'$ and $f''$ at $r=(r_{h,2D})_{lm}$ as in (\ref{enrb}) and (\ref{wcdah}), 
however these are not changed from just the Schwarzschild as can be seen from (\ref{sdghc}). 
This is the same situation with the Hawking temperature in Sec.\ref{subsec:hwkkdk} and \ref{sec:hwkttd}.

\section{Comment on  result in terms of the information paradox}  
\label{subsec:cmrst}  

As mentioned under (\ref{srfsg}), 
the result (\ref{ea0ahmn}) would be the one in the original 4D black hole, 
and if the correction is to $\varepsilon ^2$-order and $\phi$-independent,   
we could conclude by the logic in Sec.\ref{klsdlsd} 
the black-body radiation of the supertranslated black holes would be always thermal. 

Important problem for us is the information paradox. 
As an insight obtained from this work, 
the Hawking temperature and flux could not be the solution 
as we have found there is no breaking of the thermal flux 
in the range of the analysis in this paper.  

Supertranslated black hole spacetimes would be normal in reality and how to be supertranslated 
is determined by the initial configuration \cite{Compere:2016hzt}. Therefore, the information  
of the initial configuration would be stored in the configuration of the asymptotic region of the 
spacetime, again which would be the key of the information paradox.


\begin{thebibliography}{100} 
 
\bibitem{confhomp}
https://inspirehep.net/conferences/1772425 

\bibitem{Lin:2020gva} 
F.~L.~Lin and S.~Takeuchi,
``Hawking Flux from a Black Hole with Nonlinear Supertranslation Hair,''
Phys. Rev. D \textbf{102}, no.4, 044004 (2020)
[arXiv:2004.07474 [hep-th]].  

\bibitem{Bondi:1962px}    
H.~Bondi, M.~G.~J.~van der Burg and A.~W.~K.~Metzner,
``Gravitational waves in general relativity. 7. Waves from axisymmetric isolated systems,''
Proc. Roy. Soc. Lond. A \textbf{A269}, 21-52 (1962)

\bibitem{Sachs:1962wk}
R.~K.~Sachs, 
``Gravitational waves in general relativity. 8. Waves in asymptotically flat space-times,''
Proc. Roy. Soc. Lond. A \textbf{A270}, 103-126 (1962)

\bibitem{Strominger:2017zoo} 
A.~Strominger,
``Lectures on the Infrared Structure of Gravity and Gauge Theory,''
[arXiv:1703.05448 [hep-th]].

\bibitem{Compere:2016hzt} 
  G.~Compere and J.~Long,
  ``Classical static final state of collapse with supertranslation memory,''
  Class.\ Quant.\ Grav.\  {\bf 33}, no. 19, 195001 (2016) 
  [arXiv:1602.05197 [gr-qc]]. 


\bibitem{Chu:2018tzu}
C.~S.~Chu and Y.~Koyama,
``Soft Hair of Dynamical Black Hole and Hawking Radiation,''
JHEP \textbf{04}, 056 (2018)
[arXiv:1801.03658 [hep-th]].

\bibitem{Javadinazhed:2018mle}
R.~Javadinezhad, U.~Kol and M.~Porrati,
``Comments on Lorentz Transformations, Dressed Asymptotic States and Hawking Radiation,''
JHEP \textbf{01}, 089 (2019)
[arXiv:1808.02987 [hep-th]].

\bibitem{Maitra:2019eix}
M.~Maitra, D.~Maity and B.~R.~Majhi,
``Near horizon symmetries, emergence of Goldstone modes and thermality,''
Eur. Phys. J. Plus \textbf{135}, no.6, 483 (2020)
[arXiv:1906.04489 [hep-th]].


\bibitem{Wen:2021ahw}
W.~Y.~Wen,
``Dressed tunneling in soft hair,''
Phys. Lett. B \textbf{820}, 136578 (2021)
[arXiv:2103.00516 [hep-th]].

\bibitem{Barnich:2009se}
G.~Barnich and C.~Troessaert,
``Symmetries of asymptotically flat 4 dimensional spacetimes at null infinity revisited,''
Phys. Rev. Lett. \textbf{105}, 111103 (2010) 
[arXiv:0909.2617 [gr-qc]].

\bibitem{Hawking:2016msc} 
S.~W.~Hawking, M.~J.~Perry and A.~Strominger,
``Soft Hair on Black Holes,''
Phys. Rev. Lett. \textbf{116}, no.23, 231301 (2016)
[arXiv:1601.00921 [hep-th]].



\bibitem{Pshirkov:2009ak}
M.~S.~Pshirkov, D.~Baskaran and K.~A.~Postnov,
``Observing gravitational wave bursts in pulsar timing measurements,''
Mon. Not. Roy. Astron. Soc. \textbf{402}, 417 (2010)
[arXiv:0909.0742 [astro-ph.CO]].

\bibitem{vanHaasteren:2009fy}
R.~van Haasteren and Y.~Levin, 
``Gravitational-wave memory and pulsar timing arrays,''
Mon. Not. Roy. Astron. Soc. \textbf{401}, 2372 (2010)
[arXiv:0909.0954 [astro-ph.IM]].

\bibitem{NSeto}
N.~Seto,
``Search for Memory and Inspiral Gravitational Waves from Super-Massive Binary Black Holes with Pulsar Timing Arrays,''
Mon.Not.Roy.Astron.Soc. 400 (2009) L38
[arXiv:0909.1379 [astro-ph.GA]].

\bibitem{Wang:2014zls}
J.~B.~Wang, G.~Hobbs, W.~Coles, R.~M.~Shannon, X.~J.~Zhu, D.~R.~Madison, M.~Kerr, V.~Ravi, M.~J.~Keith, R.~N.~Manchester, Y.~Levin, M.~Bailes, N.~D.~R.~Bhat, S.~Burke-Spolaor, S.~Dai, S.~Oslowski, W.~van Straten, L.~Toomey, N.~Wang and L.~Wen,
``Searching for gravitational wave memory bursts with the Parkes Pulsar Timing Array,''
Mon. Not. Roy. Astron. Soc. \textbf{446}, 1657-1671 (2015)
[arXiv:1410.3323 [astro-ph.GA]].

\bibitem{Arzoumanian:2015cxr} 
Z.~Arzoumanian \textit{et al.} [NANOGrav],
``NANOGrav Constraints on Gravitational Wave Bursts with Memory,''
Astrophys. J. \textbf{810}, no.2, 150 (2015)
[arXiv:1501.05343 [astro-ph.GA]].

\bibitem{Lasky:2016knh} 
P.~D.~Lasky, E.~Thrane, Y.~Levin, J.~Blackman and Y.~Chen,
``Detecting gravitational-wave memory with LIGO: implications of GW150914,''
Phys. Rev. Lett. \textbf{117}, no.6, 061102 (2016)
[arXiv:1605.01415 [astro-ph.HE]]. 

\bibitem{Pasterski:2015tva}
S.~Pasterski, A.~Strominger and A.~Zhiboedov,
``New Gravitational Memories,''
JHEP \textbf{12}, 053 (2016)
[arXiv:1502.06120 [hep-th]].

\bibitem{Pate:2017vwa}
M.~Pate, A.~M.~Raclariu and A.~Strominger,
``Color Memory: A Yang-Mills Analog of Gravitational Wave Memory,''
Phys. Rev. Lett. \textbf{119}, no.26, 261602 (2017)
[arXiv:1707.08016 [hep-th]].

\bibitem{Bieri:2013hqa}
L.~Bieri and D.~Garfinkle,
``An electromagnetic analogue of gravitational wave memory,''
Class. Quant. Grav. \textbf{30}, 195009 (2013)
[arXiv:1307.5098 [gr-qc]]. 

\bibitem{Susskind:2015hpa}
L.~Susskind,
``Electromagnetic Memory,''
[arXiv:1507.02584 [hep-th]].
 
\bibitem{Pasterski:2015zua}
S.~Pasterski,
``Asymptotic Symmetries and Electromagnetic Memory,''
JHEP \textbf{09}, 154 (2017)
[arXiv:1505.00716 [hep-th]].

\bibitem{Strominger:2013lka}
A.~Strominger,
``Asymptotic Symmetries of Yang-Mills Theory,''
JHEP \textbf{07}, 151 (2014)  
[arXiv:1308.0589 [hep-th]].

\bibitem{He:2014cra}
T.~He, P.~Mitra, A.~P.~Porfyriadis and A.~Strominger,
``New Symmetries of Massless QED,'' 
JHEP \textbf{10}, 112 (2014)
[arXiv:1407.3789 [hep-th]].

\bibitem{Strominger:2013jfa}
A.~Strominger,
``On BMS Invariance of Gravitational Scattering,''
JHEP \textbf{07}, 152 (2014)
[arXiv:1312.2229 [hep-th]].

\bibitem{He:2014laa}
T.~He, V.~Lysov, P.~Mitra and A.~Strominger,
``BMS supertranslations and Weinberg\textquoteright{}s soft graviton theorem,''
JHEP \textbf{05}, 151 (2015)
[arXiv:1401.7026 [hep-th]].

\bibitem{Cachazo:2014fwa}
F.~Cachazo and A.~Strominger,
``Evidence for a New Soft Graviton Theorem,''
[arXiv:1404.4091 [hep-th]].

\bibitem{Kapec:2014opa}
D.~Kapec, V.~Lysov, S.~Pasterski and A.~Strominger,
``Semiclassical Virasoro symmetry of the quantum gravity $ \mathcal{S}$-matrix,''
JHEP \textbf{08}, 058 (2014)
[arXiv:1406.3312 [hep-th]].

\bibitem{Kapec:2016jld}
D.~Kapec, P.~Mitra, A.~M.~Raclariu and A.~Strominger,
``2D Stress Tensor for 4D Gravity,''
Phys. Rev. Lett. \textbf{119}, no.12, 121601 (2017)
[arXiv:1609.00282 [hep-th]].

\bibitem{He:2017fsb}
T.~He, D.~Kapec, A.~M.~Raclariu and A.~Strominger,
``Loop-Corrected Virasoro Symmetry of 4D Quantum Gravity,''
JHEP \textbf{08}, 050 (2017)
[arXiv:1701.00496 [hep-th]].

\bibitem{He:2015zea} 
T.~He, P.~Mitra and A.~Strominger,
``2D Kac-Moody Symmetry of 4D Yang-Mills Theory,''
JHEP \textbf{10}, 137 (2016)
[arXiv:1503.02663 [hep-th]].

\bibitem{Bagchi:2016bcd}
A.~Bagchi, R.~Basu, A.~Kakkar and A.~Mehra,
``Flat Holography: Aspects of the dual field theory,''
JHEP \textbf{12}, 147 (2016)
[arXiv:1609.06203 [hep-th]].

\bibitem{Pasterski:2016qvg}
S.~Pasterski, S.~H.~Shao and A.~Strominger,
``Flat Space Amplitudes and Conformal Symmetry of the Celestial Sphere,''
Phys. Rev. D \textbf{96}, no.6, 065026 (2017)
doi:10.1103/PhysRevD.96.065026
[arXiv:1701.00049 [hep-th]].

\bibitem{Cardona:2017keg}
C.~Cardona and Y.~t.~Huang,
``S-matrix singularities and CFT correlation functions,''
JHEP \textbf{08}, 133 (2017)
[arXiv:1702.03283 [hep-th]].

\bibitem{Pasterski:2017kqt}
S.~Pasterski and S.~H.~Shao,
``Conformal basis for flat space amplitudes,''
Phys. Rev. D \textbf{96}, no.6, 065022 (2017)
[arXiv:1705.01027 [hep-th]].

\bibitem{Pasterski:2017ylz}
S.~Pasterski, S.~H.~Shao and A.~Strominger, 
``Gluon Amplitudes as 2d Conformal Correlators,''
Phys. Rev. D \textbf{96}, no.8, 085006 (2017) 
[arXiv:1706.03917 [hep-th]].
 
\bibitem{Strominger:2014pwa}
A.~Strominger and A.~Zhiboedov, 
``Gravitational Memory, BMS Supertranslations and Soft Theorems,''
JHEP \textbf{01}, 086 (2016) 
[arXiv:1411.5745 [hep-th]].

\bibitem{Hawking:2016sgy} 
S.~W.~Hawking, M.~J.~Perry and A.~Strominger,
``Superrotation Charge and Supertranslation Hair on Black Holes,''
JHEP \textbf{05}, 161 (2017)
[arXiv:1611.09175 [hep-th]].

\bibitem{Strominger:2017aeh} 
A.~Strominger,
``Black Hole Information Revisited,''
[arXiv:1706.07143 [hep-th]]. 

\bibitem{Berti:2005ys}  
E.~Berti, V.~Cardoso and C.~M.~Will,
``On gravitational-wave spectroscopy of massive black holes with the space interferometer LISA,''
Phys.\ Rev.\ D {\bf 73}, 064030 (2006) 
[gr-qc/0512160].  

\bibitem{Robinson:2005pd} 
S.~P.~Robinson and F.~Wilczek,
``A Relationship between Hawking radiation and gravitational anomalies,''   
Phys.\ Rev.\ Lett.\  {\bf 95}, 011303 (2005)
[gr-qc/0502074].
  
\bibitem{Iso:2006wa}
S.~Iso, H.~Umetsu and F.~Wilczek,
``Hawking radiation from charged black holes via gauge and gravitational anomalies,''
Phys. Rev. Lett. \textbf{96}, 151302 (2006) 
[arXiv:hep-th/0602146 [hep-th]].   

\bibitem{Parikh:1999mf}

M.~K.~Parikh and F.~Wilczek,
``Hawking radiation as tunneling,''
Phys. Rev. Lett. \textbf{85}, 5042-5045 (2000)
[arXiv:hep-th/9907001 [hep-th]].


\bibitem{Lust:2018cvp}
D.~Lust and W.~Vleeshouwers, 
``Black Hole Information and Thermodynamics,''
[arXiv:1809.01403 [gr-qc]].

\bibitem{Umetsu:2010ts} 
  K.~Umetsu, 
  ``Recent Attempts in the Analysis of Black Hole Radiation,'' 
  arXiv:1003.5534 [hep-th]. 

\bibitem{wikiy3} 
https://en.wikipedia.org/wiki/Clebsch-Gordan$\_$coefficients 

\bibitem{Fujikawa:2004cx}
K.~Fujikawa and H.~Suzuki,
``Path integrals and quantum anomalies,''

\bibitem{kimura}
T.~Kimura and T.~Ohta
``Classical and quantum gravitational theory,''
McGraw-Hill Co., Inc., 1989 (Japanese), 

\bibitem{Li:2010eca} 
  R.~Li, S.~Li and J.~R.~Ren, 
  ``Hawking Radiation of Fermionic Field and Anomaly in 2+1 Dimensional Black Holes,''
  Class.\ Quant.\ Grav.\  {\bf 27}, 155011 (2010)
  [arXiv:1005.3615 [hep-th]]. 

\bibitem{Becar:2011fc}  
  R.~Becar and P.~A.~Gonzalez,
  ``Hawking Radiation for Scalar and Dirac Fields in Five Dimensional Dilatonic Black Hole via Anomalies,''
  Int.\ J.\ Mod.\ Phys.\ D {\bf 21}, 1250030 (2012)
  [arXiv:1104.0356 [gr-qc]].

\bibitem{Mao:2011zzb} 
  P.~J.~Mao, R.~Li, L.~Y.~Jia and J.~R.~Ren,
  ``Hawking radiation of Dirac particles from the Myers-Perry black hole,''
  Eur.\ Phys.\ J.\ C {\bf 71}, 1527 (2011).

\bibitem{Iso:2007nf} 
  S.~Iso, T.~Morita and H.~Umetsu,
  ``Hawking radiation via higher-spin gauge anomalies,'' 
  Phys.\ Rev.\ D {\bf 77}, 045007 (2008)
  [arXiv:0710.0456 [hep-th]]. 
  
\bibitem{Umetsu:2008cm}  
  K.~Umetsu,
  ``Ward Identities in the derivation of Hawking radiation from Anomalies,''
  Prog.\ Theor.\ Phys.\  {\bf 119}, 849 (2008)
  [arXiv:0804.0963 [hep-th]].
    
\bibitem{Banerjee:2007uc} 
  R.~Banerjee and S.~Kulkarni,
  ``Hawking radiation, effective actions and covariant boundary conditions,'' 
  Phys.\ Lett.\ B {\bf 659}, 827 (2008)
  [arXiv:0709.3916 [hep-th]].
  
\bibitem{Bertlmann:2000da} 
  R.~A.~Bertlmann and E.~Kohlprath,
  ``Two-dimensional gravitational anomalies, Schwinger terms and dispersion relations,''
  Annals Phys.\  {\bf 288}, 137 (2001)
  [hep-th/0011067].
  
\bibitem{Iso:2006ut}
S.~Iso, H.~Umetsu and F.~Wilczek,
``Anomalies, Hawking radiations and regularity in rotating black holes,''
Phys. Rev. D \textbf{74}, 044017 (2006)
[arXiv:hep-th/0606018 [hep-th]]. 

\bibitem{Murata:2006pt} 
  K.~Murata and J.~Soda,
  ``Hawking radiation from rotating black holes and gravitational anomalies,''
  Phys.\ Rev.\ D {\bf 74}, 044018 (2006)
  [hep-th/0606069].
  
\end{thebibliography}
\end{document}